\documentclass[aps,prb,twocolumn,amsmath,amssymb,gallery,footinbib,superscriptaddress,floatfix]{revtex4-2}
\usepackage{graphicx}
\usepackage{dcolumn}
\usepackage{bm}
\usepackage{hyperref}
\usepackage{stackrel,amssymb}
\hypersetup{colorlinks=true, linkcolor=blue, citecolor=blue, urlcolor=magenta, pdftitle={On the search of displacive chiral phase transitions}, pdfauthor={F. Gomez-Ortiz}}
\usepackage{bbding}
\usepackage{xcolor} 
\usepackage[normalem]{ulem}

\begin{document}
\preprint{APS/123-QED}
\title{Role of long-range dipolar interactions in the simulation of the properties of polar crystals using effective atomic potentials}
\author{Miao Yu}
\affiliation{Theoretical Materials Physics, Q-MAT, Université de Liège, B-4000 Sart-Tilman, Belgium}
\affiliation{Institute of Atomic and Molecular Physics, Sichuan University, Chengdu 610065, China}
\author{Fernando G\'omez-Ortiz}
\affiliation{Theoretical Materials Physics, Q-MAT, Université de Liège, B-4000 Sart-Tilman, Belgium}
\author{Louis Bastogne}
\affiliation{Theoretical Materials Physics, Q-MAT, Université de Liège, B-4000 Sart-Tilman, Belgium}
\author{Jin-Zhu Zhao}
\affiliation{Guangdong Provincial Key Laboratory of Quantum Engineering and Quantum Materials, School of Physics, South China Normal University, Guangzhou 510006, P. R. China}
\affiliation{Guangdong-Hong Kong Joint Laboratory of Quantum Matter, South China Normal University, Guangzhou 510006, P. R. China}
\author{Philippe Ghosez}
\email[Corresponding author: ]{Philippe.ghosez@uliege.be}
\affiliation{Theoretical Materials Physics, Q-MAT, Université de Liège, B-4000 Sart-Tilman, Belgium} 

\date{\today}
\begin{abstract}
Driven by novel approaches and computational techniques, second-principles atomic potentials are nowadays at the forefront of computational materials science, enabling large-scale simulations of material properties with near-first-principles accuracy.
However, their application to polar materials can be challenging, particularly when longitudinal-optical phonon modes are active on the material, as accurately modeling such systems requires incorporating the long-range part of the dipole-dipole interactions.
In this study, we challenge the influence of these interactions on the properties of polar materials taking BaTiO$_3$ as paradigmatic example. By comparing models with and without the long-range part of the electrostatic contributions in a systematic way, we demonstrate that even if these interactions are neglected, the models can still provide an overall good description of the material, though they may lead to punctual significant artifacts.
Our results propose a pathway to identify when an atomistic potential may be inadequate and needs to be corrected through the inclusion of the long-range part of dipolar interactions.
\end{abstract}
\maketitle
\section{Introduction}
\label{sec:introduction}
For decades, empirical and semi-empirical models have been fundamental in the atomistic modeling of materials, providing an alternative framework even prior to first-principles methods. Early approaches relied on interatomic potentials such as Lennard-Jones~\cite{Jones-24}, Stillinger-Weber~\cite{Weber-85}, and Tersoff~\cite{Tersoff-88} potentials, which successfully described properties of simple materials.
Rapidly, they improved and became more sophisticated~\cite{Pedone2006} and today a wide range of strategies, depending on the specific material system, can be found in the literature~\cite{Liang-13,Harrison-18} providing highly accurate predictions of material properties. 
With advances in computing power and electronic‑structure theory~\cite{Kohn-65,Hohenberg-65,Perdew-96,Payne-92}, the community increasingly turned to first‑principles simulations for greater predictive accuracy. 
However, aided by modern parameter fitting algorithms, high-quality reference data, and new perspectives, the atomistic modeling of materials is now enjoying a resurgence, allowing simulations of longer length and time scales as well as including operating conditions, such as the application of external electric fields or mechanical constraints~\cite{Ghosez-22}. 
Atomistic models directly fitted on first-principles data (and sometimes referred to as ``second-principles") now offer a versatile framework that allows researchers to create custom models for a wide range of materials ranging from metals and semiconductors to insulators~\cite{Behler-07,Chmiela-17,Jinnouchi-19,Grosse-25}.
Modern approaches include pioneering lattice effective Hamiltonians~\cite{Zhong-94,Rabe-95}, related effective atomic potentials~\cite{Wojdel-13,Escorihuela-17,García-Fernández-16,Gonze-20}, phase-fields methodologies~\cite{Long-Qing}, shell- and bond-valence-models~\cite{Shell,Bondv}, reactive force fields~\cite{vanDuin-01} and their application to ferroelectrics~\cite{Ganesh-19,Ganesh-22,Ganesh-23} or recent machine-learning interatomic potentials, including emerging universal potentials~\cite{Universal}. 

Recently, with the advent of machine learning, atomic potentials have indeed experienced a revolution, and are rapidly becoming a standard complementary tool in computational materials science~\cite{Behler-07, Chmiela-17, Zhang-18, Botu-15, Jinnouchi-19, Grosse-25} to extend the predictive power of first-principles simulations in a multiscale context. Approaches such as Gaussian process regression~\cite{Klawohn-23}, kernel methods~\cite{Botu-15} and neural networks~\cite{Zhang-18,Singraber-19} have democratized the generation of accurate interatomic models for many different materials.
These models are typically trained on extensive datasets generated from first-principles calculations without the inclusion of any extra terms and relying on the configuration itself rather than distortions with respect to a high-symmetry reference structure.
As a result, long-range interactions are often neglected. Although these models are typically assumed to depend on the local atomic environment at first sight, the actual range of the interactions is dependent and inherently determined by the size of the training set used to develop the model. Therefore, the correct description of the nonanalytic behavior of the interatomic force constants in the long-wavelength limit presents a challenge as it originates from the power-law behavior of the Coulombic interaction and cannot be easily captured by spatially truncated training sets. 
Moreover, the absence of a reference structure inherently complicates the analytical inclusion of electrostatic interactions, which are often neglected in AI models beyond the spatial range covered by their training set.
Novel approaches are trying to circumvent this issue including charge equilibration schemes~\cite{Shaidu-24}, the on-the-fly prediction of maximally localized Wannier function centers~\cite{Zhang-22,Gao-22} or a recently proposed modification of the model without the need for additional first-principles calculations or retraining discussed in Ref.~\cite{Monacelli-24}. However, these tools have not yet become common practice, and most of the machine-learned interatomic potentials rely on the nearsightedness of interatomic interactions neglecting such long range part of the interactions beyond the range defined by the training set which in most cases turns out to be computed on a 2$\times$2$\times2$ supercell.

In this work, we aim to clarify the concrete impact of neglecting the long-range part of the dipole-dipole interactions in computational models and to identify the scenarios where incorporating corrections for an accurate description becomes crucial. By focusing on the prototypical ferroelectric material BaTiO$_3$ in which dipolar interactions are expected to be key ~\cite{Cochran-60, Ghosez-96}, we systematically explore the influence of the dipole-dipole interactions on the phonon dispersion curves and atomic distortion patterns. 
On the one hand, we demonstrate that the inclusion of the long-range part of dipolar interactions is not always essential for an overall accurate description of the system, including the ferroelectric instability and its characteristic chain-like character, which supports the good performance of AI-based models even in polar materials~\cite{Gigli-22}.
On the other hand, we also point out specific cases where neglecting the long-range part of the dipole-dipole interaction can lead to significant artifacts in the phonon spectrum, resulting in inaccurate descriptions of the energy landscape associated with certain atomic displacements and the emergence of spurious phases.

\section{Methodology}
\label{sec:Methods}
To assess the impact of the long-range character of dipolar interactions on the lattice dynamics of BaTiO$_3$, we rely here on a second-principles effective atomic potential (EAP) method, as implemented in the {\sc{multibinit}} software. The method relies on a Taylor expansion of the Born-Oppenheimer potential energy surface (PES) in terms of structural degrees of freedom around the cubic perovskite structure taken as reference. 
Although lower symmetry phases like the $P4mm$, $Amm2$, or $R3m$ could also be considered as proper references for constructing the model, the choice of the  cubic  parent phase follows previous similar studies in perovskites based on Taylor expansion \cite{Zhong-94, Wojdel-13, Ghosez-22} and is the most sensible choice to access the whole phase diagram. 
The Taylor expansion around the reference geometry is made in terms of individual atomic displacements and homogeneous strain degrees of freedom and includes phonon, strain and strain-phonon energy terms at both harmonic and anharmonic levels. The coefficients of harmonic terms corresponds to second-energy derivatives which are straightforwardly determined from density-functional perturbation theory, so that the model keeps first-principles accuracy at the harmonic level. Then, most relevant anharmonic terms, which capture key deviations from harmonic behavior, are selected and their coefficients fitted in order to reproduce the energies, forces, and stresses as obtained from DFT calculations for a representative set of atomic configurations. The reference model used here is a slight revision of the one reported in Ref.~\cite{Gomez-25}, only including refitted anharmonic terms. Parameters and a validity assessment passport are provided in Appendix.  

In {\sc{multibinit}}, the dipole-dipole contribution to the interatomic force constant (IFC), $C^{DD}_{\kappa \alpha, \kappa' \beta}$, related to atomic displacement of atom $\kappa$ in direction $\alpha$ and atom $\kappa'$ in direction $\beta$, separated by a vector ${\bf d}$ are by default evaluated analytically from the knowledge of Born effective charges $Z^*$ and optical dielectric tensor $\epsilon_{\infty}$ using an expression that generalize the following classical formula, 
\begin{equation}
C^{DD}_{\kappa \alpha, \kappa' \beta}
=\frac{Z^*_{\kappa} Z^*_{\kappa'}}{\epsilon_{\infty}}
(\frac{\delta_{\alpha \beta}}{d^3}-3 \frac{d_{\alpha} d_{\beta}}{d^5}),
\end{equation}
to anisotropic cases following the scheme described in Ref. ~\cite{Gonze-97}.

\begin{figure}[tbhp]
     \centering
      \includegraphics[width=6cm]{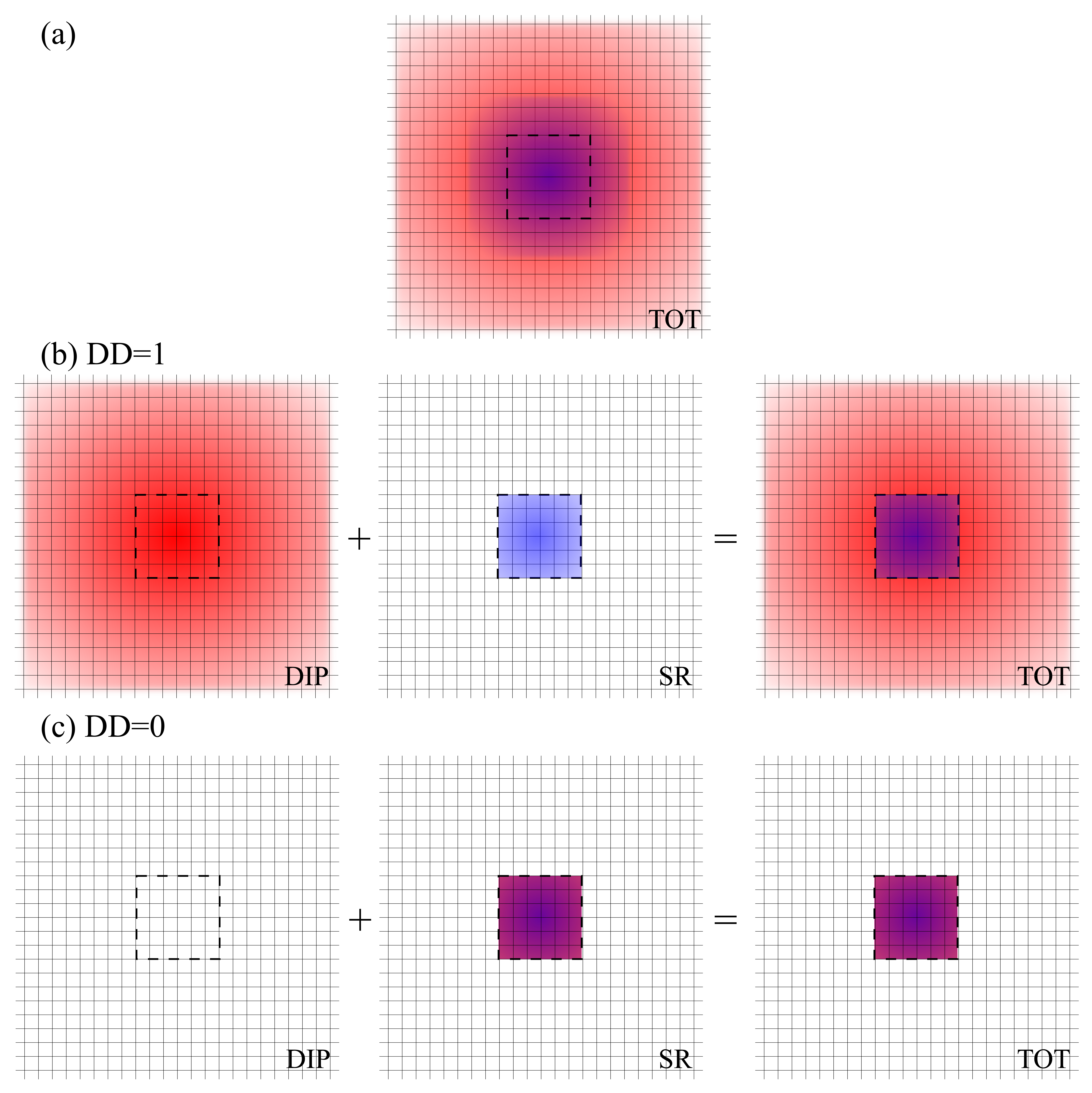}
      \caption{(a) Schematic representation of the  dipole-dipole (DD, red) and short-range (SR, blue) contributions to the total interatomic force constants (IFC). (b) When activating the explicit treatment of DD interactions, the latter are computed analytically and infinite range while the SR IFC are obtained by subtracting the DD part from the total IFC and their range limited to that of a supercell defined by the $q$-point grid used for the calculations. (c) When deactivating the explicit treatment of DD interactions, the SR IFC correspond to the total IFC but truncated at the range of the supercell defined by the $q$-point grid. }
      \label{fig:interactions} 
\end{figure}

In practice, and as illustrated in Fig. \ref{fig:interactions}, this analytical dipole-dipole (DD) contribution -- which extents from nearest neighbors to infinite range -- is subtracted from the total IFC to yield the remaining so-called short-range (SR) IFC. The process is associated to a Fourier transform of the dynamical matrices on a $n \times m\times p$ grid of $q$-points, which practically limits the SR IFC to a range defined by a corresponding  $n \times m \times p$ supercell. Inside the range of this supercell, the total IFC are therefore the sum of SR and DD contributions, while outside it restricts to the DD contribution. Alternatively, the explicit treatment of DD interactions can also be set to zero, in which case the SR IFC correspond to the total IFC but with a range artificially cut to the size of the supercell. 

\begin{figure}[tbhp]
     \centering
      \includegraphics[width=\columnwidth]{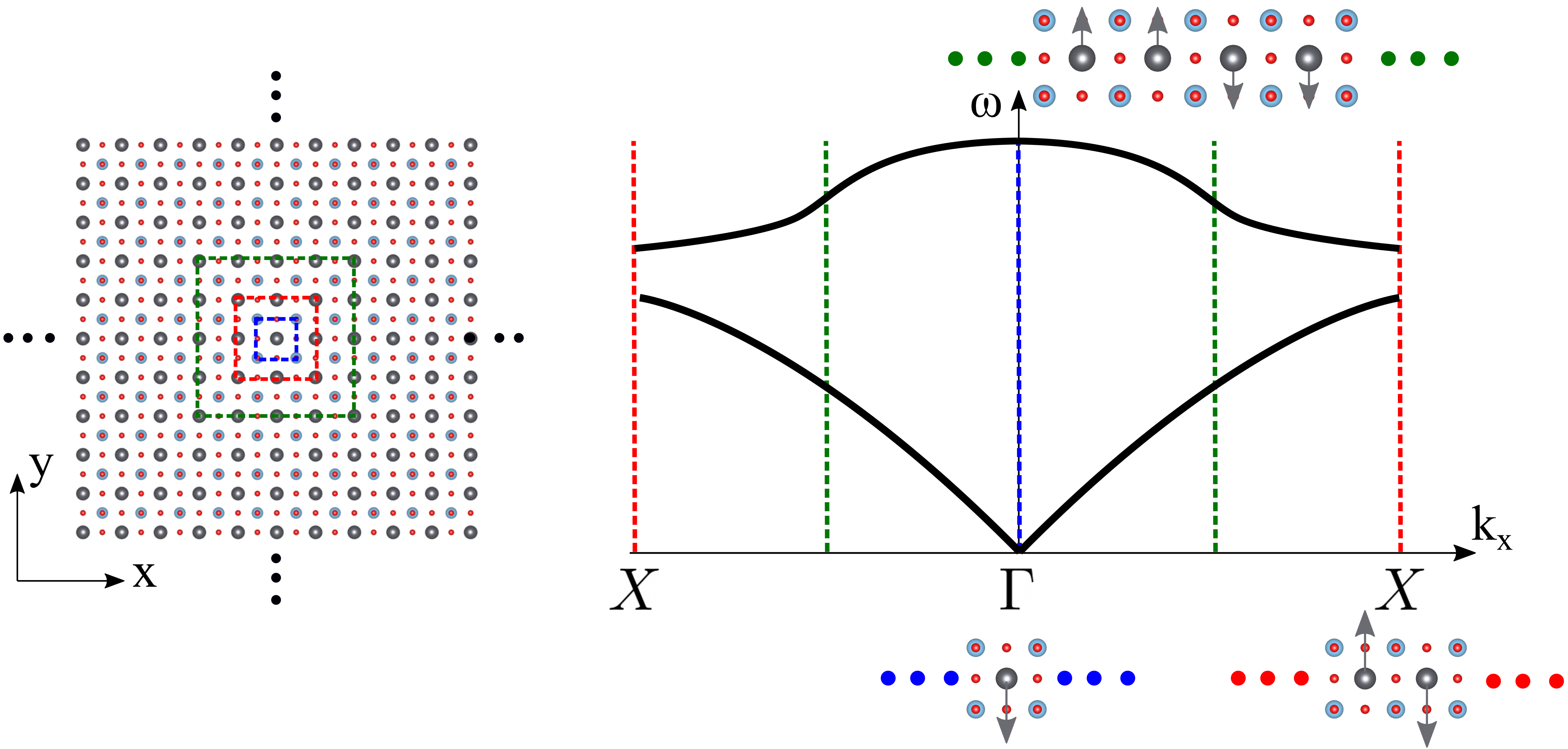}
      \caption{Illustration of the relationship between $1\times1\times1$ (blue), $2\times2\times2$ (red) and $4\times4\times4$ (green) real-space supercells (projected on the $x$-$y$ plane) and reciprocal-space $q$-point grids (along $k_x$). The real-space atomic distortion patterns allowed in a given supercell are restricted to the $q$-points of the associated reciprocal-space grid. due to the periodic boundary conditions, simulations within a given supercell only provide access to interatomic force constants up to the range defined by the supercell. On the other hand,  atomic displacements compatible with a given supercell are restricted to linear combination of the eigenvectors of the phonon modes associated to the associated $q$-point grid.   
      }
      \label{fig:scheme} 
\end{figure}

Fig. \ref{fig:scheme} illustrates the correspondence between the size of the supercell used in real space and that of the related $q$-point grid in reciprocal space. The atomic distortion patterns allowed in a given $n \times n \times n$ supercell are linear combinations of the eigendisplacements vectors of the phonon modes associated to the corresponding $n \times n \times n$ grid of $q$-points. Looking at the frequencies of the phonons associated to that mesh (and comparing model to DFT data) so quantifies the energetics, at harmonic level, of the displacement patterns compatible with that supercell. Looking at intermediate points belonging to finer meshes, allow to estimate the energetics of patterns in bigger supercells. 

Second-principles models are typically trained on first-principles data using a given supercell. Looking at how the phonons on the related grid are reproduced by the model allow to assess the precision of the model at harmonic level. In the case of {\sc{multibinit}}, the model is exact by construction at that level. Then the model is used for simulations in larger supercells. Looking at the quality of the dispersion curves at intermediate points related to finest grids allow to anticipate the precision of the model in such bigger supercells.

By simultaneously adjusting the real‑space cutoff used to compute the interatomic force constants and toggling the analytical dipole–dipole term in {\sc multibinit}, one can systematically modulate the effective interaction range. This dual control offered by {\sc multibinit} provides a unique way to progressively refine models by incorporating interactions at larger ranges, enabling a controlled analysis of their impact through simulations on different supercell sizes. Moreover, since non-dipolar short-range interactions are expected to decay exponentially fast according to the nearsightedness principle, this approach allows us to assess the specific impact of the dipole–dipole interaction range, as will be discussed in the Results section.

The different models discussed on this article are constructed using different $q$-point meshes for the calculation of the dynamical matrices. 
\begin{figure*}[th]
     \centering
      \includegraphics[width=\textwidth]{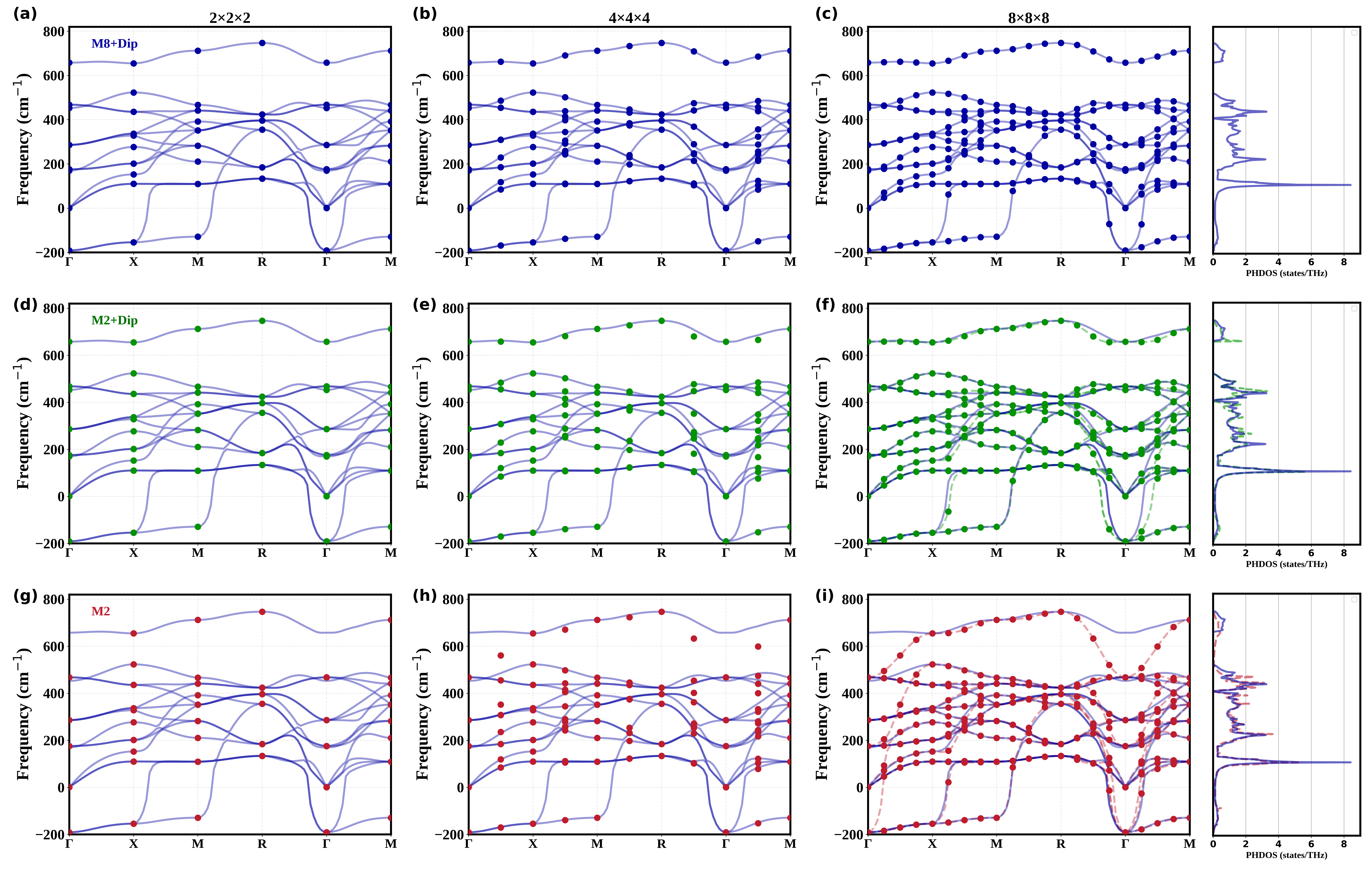}
      \caption{Phonon dispersion curves of BaTiO$_3$ obtained from three distinct models each capturing a specific range of interactions (8, 2 and 2 from top to bottom rows) with varying levels of resolution in the phonon band structure (2, 4 and 8 from left to right columns). The first two models (blue, green) include analytically the dipole-dipole interactions that extend to infinite range \texttt{M8+Dip} and \texttt{M2+Dip}, while the third (red) omits them \texttt{M2}. The solid blue line present in all panels represents the reference model (8$\times$8$\times$8 $q$-point mesh with analytic treatment of the dipolar interactions).}
      \label{fig:phonons} 
\end{figure*}
The first model (blue points in Fig.~\ref{fig:phonons}) is derived using an 8×8×8 $q$-point mesh, while the second (green points) and third (red points) models with a 2×2×2 mesh. 
Notably, with the described approach, the dipole-dipole interactions are inherently included up to the specified interaction ranges (8, and 2 unit cells respectively).
%
Some of the models, denoted as ``\texttt{+Dip}" will include the analytical dipole-dipole correction at infinite range, while others do not, limiting consequently the dipole-dipole interactions to the specific range determined by their respective $q$-mesh as schematized in Fig.~\ref{fig:interactions}, allowing us to systematically assess its influence.
Interestingly, since the anharmonic interactions in the model are limited to a range of $\sqrt{3}/2$ times the lattice constant on the reference structure, and no differences at the harmonic level are expected for cutoffs below two unit cells, all models effectively share the same anharmonic description—rendering them equivalent in this respect.

Remarkably, the model derived with a 2×2×2 $q$-point mesh, which excludes the analytical dipole-dipole correction, is of special interest, as it mimics the typical interaction range used in the training sets of AI models, which in turn defines the maximum range of interactions these models incorporate. At the other extreme, the model with 8×8×8 $q$-point mesh, including the analytical dipole-dipole correction can be considered as the exact model to be taken as reference.

Structural relaxations were performed using the Broyden-Fletcher-Goldfarb-Shanno (BFGS) method~\cite{fletcher2000practical} and finite temperature simulations followed a Hybrid molecular dynamics Monte-Carlo approach (HMC)~\cite{duane1987hybrid,betancourt2017conceptual}, consisting of Markov chain Monte Carlo sampling.

\section{Results}
Let us first explore the influence of neglecting the long-range part of the dipole dipole interactions in the phonon dispersion curves of BaTiO$_3$. 
Figure~\ref{fig:phonons} presents the phonon dispersion curves produced by three distinct models capturing each a given range of interactions  (arranged by rows) at different levels of resolution in the phonon band structure (arranged by columns). The latter intent to assess the quality of the model by quantifying in reciprocal space the energetics of distortion patterns compatible with supercells of increasing sizes, not necessarily limited to the intrinsic range of interactions included in the model. 

In the first column, a 2×2×2 supercell is considered, which only give access to distortions patterns associated to  phonon modes at the high-symmetry points (dots at $\Gamma,~X,~$...) as schematized in Fig.~\ref{fig:scheme}. The second column (4×4×4 supercell) extend this to additional intermediate points along the dispersion paths. Finally, the third column (8×8×8 supercell) further increases the number of points offering a more detailed sampling of the phonon spectrum and overlays a faint colored line (blue, green and red) representing the frequencies as predicted by the model in the infinite supercell limit.

By rows, each model is constructed, as explained in Sec.~\ref{sec:Methods}, using different $q$-point meshes for the calculation of the dynamical matrices and related to a specific  range of interatomic-force constant in real space. The first-row model (blue points) is derived using an 8$\times$8$\times$8 $q$-point mesh, while the second-row model (green points) and the third-row model (red points) with a 2$\times$2$\times$2 mesh. 
It is important to emphasize that, within the described framework, the short-range part of the dipole-dipole interactions is inherently included up to the specified interaction range of each model.
The long-range part of the dipole-dipole interactions is treated differently among the models. In the first two models, it is incorporated analytically as discussed in Sec.~\ref{sec:Methods}. In contrast, the third model omits this contribution. We will refer to them as \texttt{M8+Dip}, \texttt{M2+Dip}, and \texttt{M2}, respectively.

For reference, the solid blue line, representing the phonon frequencies predicted by the \texttt{M8+Dip} model constructed with an 8×8×8 $q$-point mesh and including long-range part of the dipole-dipole interactions, is superimposed on all panels.
Importantly, the long-range part of the dipole-dipole interactions is only included in the plot of the phonon dispersion lines if the model accounts for this information. 

We first compare the results of the 2$\times$2$\times$2 model with the analytic electrostatic interactions (\texttt{M2+Dip}) with those of the reference model \texttt{M8+Dip} (8x8x8 $q$-mesh and full dipole-dipole contributions). As shown in Fig.~\ref{fig:phonons}, we notice only minor differences in the phonon dispersion curves, which show good agreement with what is reported in the literature~\cite{Ghosez-98}. This suggests that beyond a range of two unit cells, the interactions are mostly dominated by electrostatics while the short-range part nearly vanishes~\cite{Chen-24}. This observation is in line with the results of Ref.~\cite{Ghosez-98} where the inclusion of the point $\Lambda=(0.25,0.25,0.25)$ on top of the 2$\times$2$\times$2 $q$-point mesh was anticipated to provide good convergence of the phonon dispersion curves.

We then investigate the difference between the phonon dispersion curves obtained using the model constructed with a 2$\times$2$\times$2 $q$-point mesh without the long-range component of dipole-dipole interactions (\texttt{M2}) and those obtained  with the models that include this component \texttt{M8+Dip} and \texttt{M2+Dip}. When this long-range part of the dipolar contribution is omitted, Fig.~\ref{fig:phonons} clearly shows that while the overall description of the phonon dispersion bands remains accurate, certain specific bands, particularly near the $\Gamma$ point, are inadequately described.
This is evidenced by the extra unstable LO branch and the artificial degeneracy between the LO and TO modes of the topmost band at $\Gamma$. These discrepancies underscore the critical role of the long-range part of electrostatic interactions in accurately capturing the giant LO-TO splitting characteristic of BaTiO$_3$~\cite{Zhong-94.2,Ghosez-97}.
Including the long-range part of the dipole-dipole interactions corrects these inaccuracies, yielding a significantly improved representation of the phonon bands in this region. Nevertheless, the model is in good agreement with the reference dispersion curves for the remaining bands and for the LO branches away from $\Gamma$, as indicated by the overlap of the red curves with the blue one in Fig.~\ref{fig:phonons}(f,i).

Interestingly, the ferroelectric instability, intrinsic to TO modes, is very well captured even without incorporating the long-range part of electrostatic interactions in the model. This demonstrates that the short-range part of the Coulomb interactions alone (limited to two unit cells in this case) are sufficient to accurately describe this instability, indicating that in BaTiO$_3$, the long-range part of the Coulomb interaction does not play a significant role in its ferroelectric character.
This may sound surprising at first sight since the ferroelectric instability has historically been explained from the Cochran picture~\cite{Cochran-60} as the competition between the short-range and the dipole-dipole interactions. This picture was validated by first principles calculations in Ref.~\cite{Ghosez-96}. However, it is important to note that, as we have already mentioned before, the dipole-dipole interactions include both short-range and long-range parts and the competition between the so-called short-range and dipole-dipole interactions is already present in the interatomic force constants down to first neighbors as studied in Ref.~\cite{Ghosez-99}. Our results simply highlight this fact and demonstrate that, in BaTiO$_3$, the dominant contributions of the dipole-dipole interactions responsible for the ferroelectric instability are already well captured within a 2$\times$2$\times$2 supercell. 

\begin{figure}[tbhp]
     \centering
      \includegraphics[width=\columnwidth]{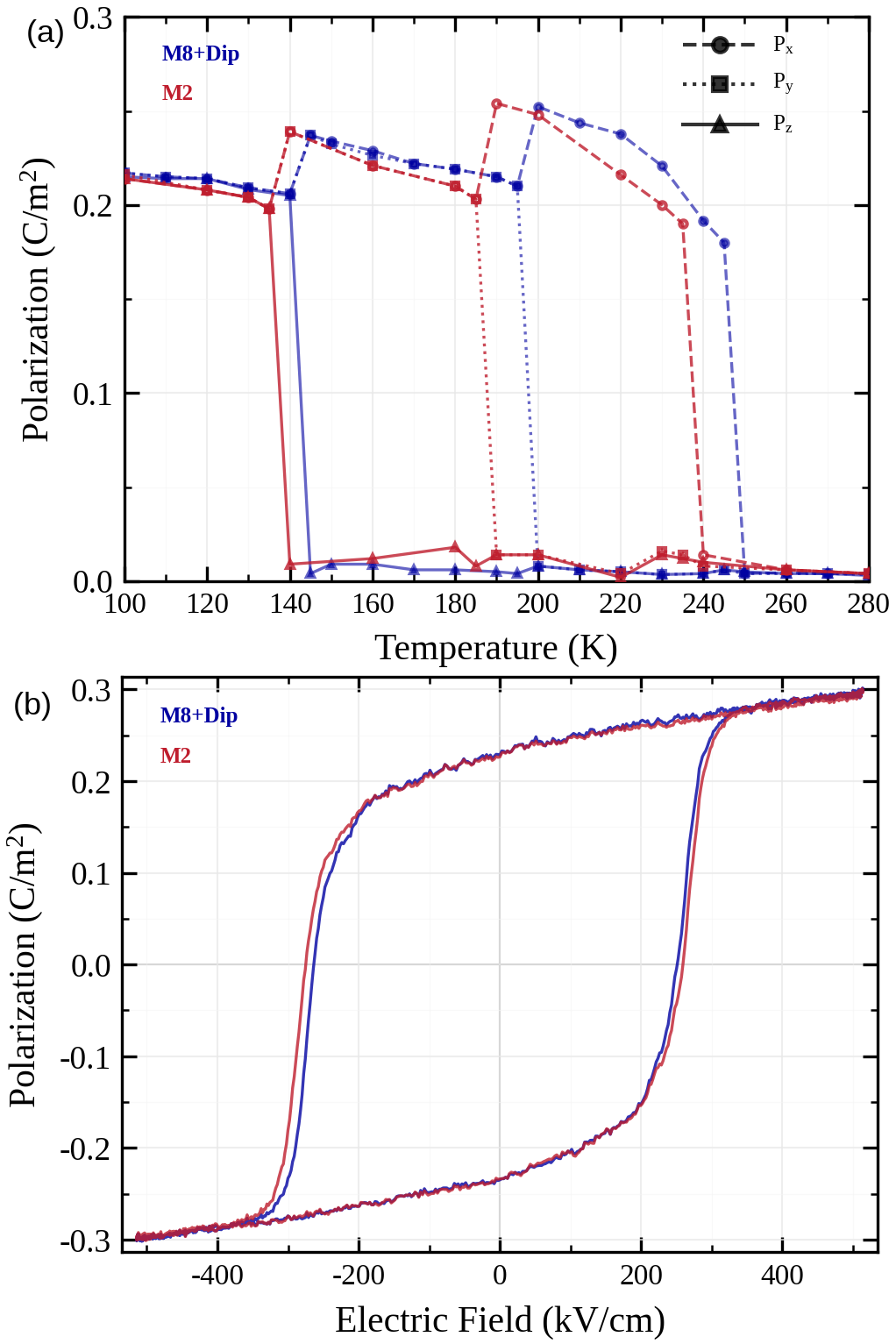}
      \caption{(a) Phase diagram of BaTiO$_3$, and (b) hysteresis loop of the $R3m$ phase at 50 K, reproduced using the \texttt{M2} and \texttt{M8+Dip} models, as indicated in the legend. The supercell size used for the calculations was 12$\times$12$\times$12.}
      \label{fig:hyst} 
\end{figure}

This establishes for instance, that if a BaTiO$_3$ model is built using AI methods with a 2$\times$2$\times$2 supercell for the training set, and so limiting implicitly the interaction range to 2 unit cells (equivalently to our 2$\times$2$\times$2 model without dipole-dipole), the overall description of the material will remain largely accurate (the ferroelectric instability and its chain-like character \cite{Ghosez-98} are preserved), except for the specific LO phonon branches affected by the missing long-range part of dipolar interactions, particularly near $\Gamma$. This discrepancy is unlikely to significantly impact the material's dynamics and phase transitions, as the phonon density of states remains well reproduced and the number of poorly described phonon bands is marginal.
This is indeed confirmed in Fig.~\ref{fig:hyst} where it appears that the phase transition sequence and hysteresis loops reproduced by \texttt{M2} and \texttt{M8+Dip} models are very closely similar which explains why AI-based models limited to short-range interactions have been shown to successfully predict phase diagrams and ferroelectric transitions in complex materials~\cite{Gigli-22,Gigli-24,Zhang2025,Thong-23,Feng-25}.
However, as it will be emphasized later, the improper treatment of the long-range part of dipolar interactions may introduce artificial discrepancies in the energy landscape, potentially affecting specific analyses.
\begin{figure*}[tbhp]
     \centering
      \includegraphics[width=\textwidth]{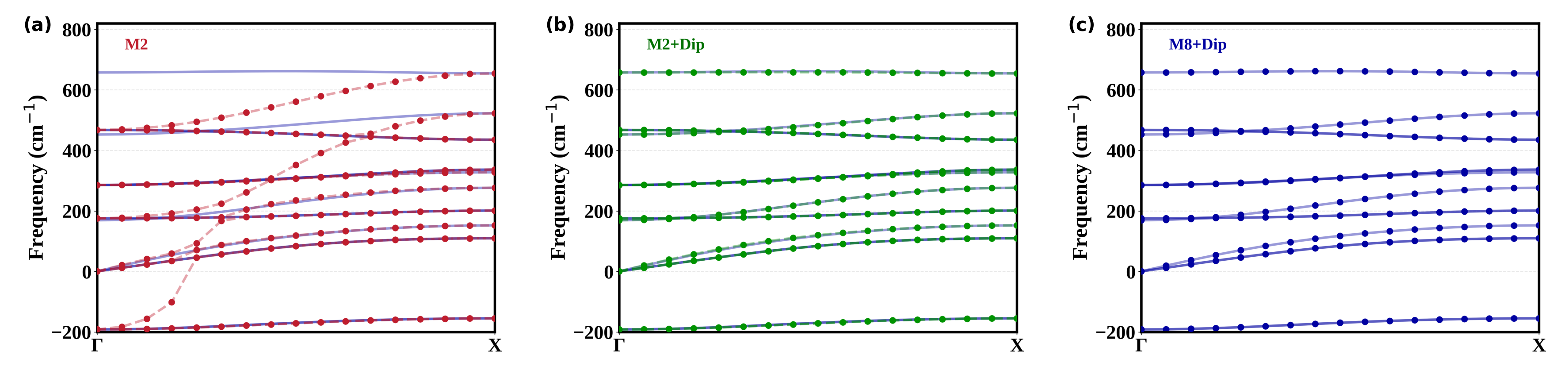}
      \caption{Phonon dispersion curves of BaTiO$_3$ along the $\Gamma-X$ branch for the same three models as described in Fig.~\ref{fig:phonons}. Blue lines corresponds to the phonon dispersion curves computed with the reference model.}
      \label{fig:gammaX} 
\end{figure*}
It must also be noted  that, although the present discussion might have some general character, the specific 2$\times$2$\times$2 range is a priori material dependent and cannot be taken for granted in a general case.

It is important at this stage to clarify the distinction between the plotted phonon dispersion curves and the information within the model. Programs like \textsc{Phonopy}~\cite{Togo-23} can include the contribution of dipole-dipole interactions on top of the information provided by the \texttt{M2} model, thereby mimicking the phonon dispersion curves of \texttt{M2+Dip} model plotted in Fig.\ref{fig:phonons}(f). 
This is indeed a common practice in the literature~\cite{Thong-23,Zhang-24,Gigli-22}. However, this correction is artificially made at the level of the post-processing while it is not present in the independent use of the model creating the illusion that the model phonons correspond to those plotted on Fig.\ref{fig:phonons}(f), while in reality, the model remains at the level of Fig.~\ref{fig:phonons}(i). 

To conduct a deeper analysis, we now focus on the $\Gamma-X$ branch of the phonon dispersion curve, which allows us to explore larger supercells. The results, presented in Fig.~\ref{fig:gammaX}, demonstrate that certain features, such as an extra unstable phonon branch predicted by the \texttt{M2} model, are artifacts arising from the absence of the long-range part of the dipole-dipole interactions. When these interactions are included, the branch stabilizes. Similarly, the high-frequency LO mode is significantly corrected, eliminating its near degeneracy with the TO mode near the $\Gamma$ point, as predicted by the model when the long-range part of dipolar interactions are not included.
In contrast, the remaining bands are reasonably well described.
\begin{figure*}[tbp]
    \centering
    \includegraphics[width=15.5cm]{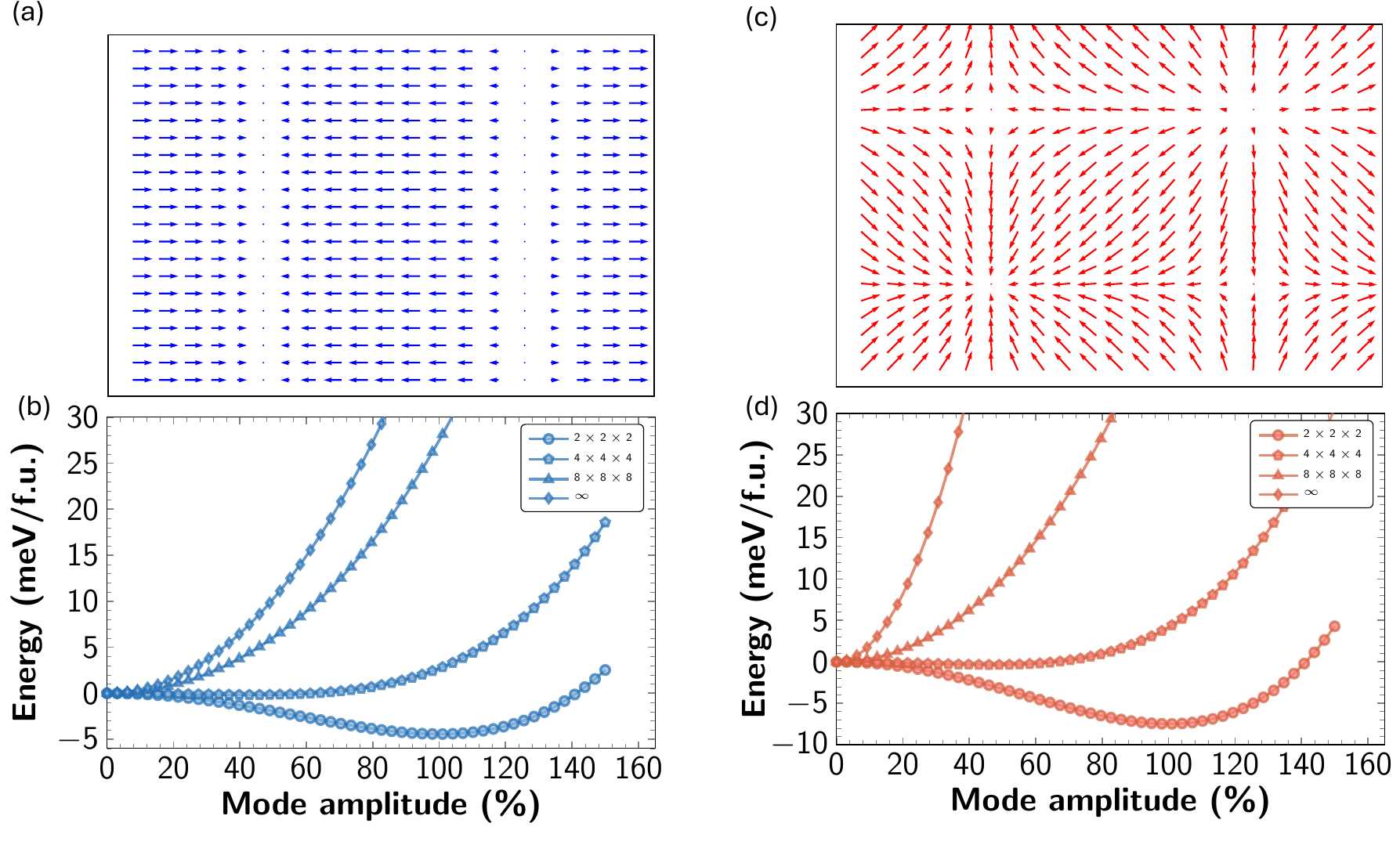}
    \caption{Polarization map resulting from the condensation of (a) one and (c) two LO phonon modes (directed along $y$ and $x$) from the $\Gamma-X$ branch at $q=1/20$ u.c.$^{-1}$ predicted by the model calculated using a 2x2x2 $q$-point mesh and excluding the long range part of the dipole dipole interactions. (b,d) Energy landscapes as a function of the amplitude of the corresponding LO modes predicted by the models calculated using a 2x2x2, 4x4x4, 8x8x8 $q$-point  mesh without the long range part and a 2x2x2 with the long range part of the dipole-dipole interactions as specified in the legend.}
    \label{fig:polphase}
\end{figure*}

Some key conclusions can be drawn at this stage.  First, neglecting the long-range part of the dipole-dipole interactions leads to significant inaccuracies in the description of the LO modes, making it essential to include them whenever these bands are involved in the materials distortions. Second, accounting for these interactions not only improves the accuracy of the phonon frequencies but also speeds up convergence, allowing a good description of the phonon dispersion curves with a relatively crude $q$-point grid value and providing a clear computational advantage as the short-range interactions are already reasonably well described with a supercell of size 2. Finally, a relatively small interaction cutoff  might be sufficient to achieve a reasonable global description of the phonon dispersion curves for most materials (besides the description of the LO modes), even capturing the ferroelectric instability. This explains the good performance of AI methods in accurately describing the phase diagram of BaTiO$_3$ and other compounds even when trained on relatively small supercells~\cite{Gigli-22}. However, it is important to keep in mind that this range is material dependent.  

To point out the main and concrete deficiencies arising from the exclusion of the long-range component of the dipole-dipole interactions, we can condense the LO phonon at $q=1/20$ u.c.$^{-1}$ from the spurious unstable branch predicted by the \texttt{M2} model  (Fig. \ref{fig:gammaX}a)  
As shown in Fig.~\ref{fig:polphase}, the condensation of such LO modes induce head-to-head and tail-to-tail domains that remain metastable due to the wrong description of the phonon branch. Clearly, this polarization state is unphysical as the divergence of the polarization generates bound charges on the material that enormously increase the electrostatic energy of the system. Similarly, when condensing together two of these LO phonon modes directed along $x$ and $y$, we can stabilize a lattice of center-convergent and center-divergent vortices demonstrating that more complex polar phases can also suffer from this spurious behavior.
As shown in Fig.~\ref{fig:polphase}, the energy lowering associated to such fake metastable phase is not negligible ($\approx 10 $ meV/f.u.) and comparable to that of the $Pmma$ phase ($\approx 10 $ meV/f.u.). 

Properly including the long-range part of the dipole-dipole interactions, as discussed earlier, stabilizes the problematic phonon LO branch and corrects the issue, leading to the suppression of the related fake low-energy states. As shown in Fig.~\ref{fig:polphase}(c), the energy landscape as a function of the phonon mode amplitude changes dramatically with the inclusion of the long-range part of the dipole-dipole interactions. The original double-well potential is replaced by a steeply rising energy surface, preventing the formation of bound charges and restoring a proper description of the material.

In the present examples, the apparent minima observed in Fig.~\ref{fig:polphase}(a) and (c) when limiting the range of interactions, are in fact saddle points of the Born-Oppenheimer energy surface once full structural relaxation without symmetry constraints is allowed. This is confirmed by finite-temperature simulations using the M2 models, where the associated atomic configurations are unstable and do not persist. This result is consistent with the absence of report of such metastable phases in previous AI models restricted to short-range interactions. But we cannot exclude the presence of eventual spurious local minima in other cases and, at least, our Fig.~\ref{fig:polphase} demonstrates that models limited to short-range interactions will significantly over-stabilize head-to-head or tail-to-tail polar configurations, which will unavoidably affect phase competition.

In short, it is important to know that as soon as the longitudinal optical phonon modes are involved in the atomic displacements of the material under study, the proper treatment of the long-range part of the dipole-dipole interactions remains crucial to properly quantify the energetics. This might become relevant when studying topological textures, dynamics of large supercells or thin films and superlattices where depolarizing fields play an important role.

\section{Conclusions}
In summary, our results demonstrate that the overall phonon dispersion curves as well as structural phase transitions with temperature and hysteresis loops  are well captured even when the interactions are limited to a relatively small range which explains the success of AI-based models in describing material properties, where the range of interactions is inherently limited by the supercell size used to generate the training set. Particularly surprising is the fact that the ferroelectric instability is very well described even in the case of BaTiO$_3$, where long-range electrostatic interactions were thought to be crucial for its ferroelectric character. Our results show, however, that while these dipolar interactions are crucial, it is the short-range part of them that appears to play a key role in driving this effect.

However, despite these successes, some problems can arise as a consequence of neglecting the long-range part of the Coulomb interactions. In particular, some artifacts such as fake minima and improper treatment of LO modes, which lead to inaccuracies in the phonon spectrum at particular bands and in the energetics of related distortion patterns. These issues suggest that further refinement of the AI models is necessary whenever atomic distortions associated to LO modes are active on the material under study. 

On the light of these results, a useful strategy for determining when long-range electrostatic corrections are needed to properly predict the energetics of some distortion patterns is to project the atomic displacements onto the phonon modes and check how much the LO modes contribute. 
This approach helps to identify \textit{a posteriori} when the model’s accuracy might be compromised, allowing for necessary refinements to improve its predictive power. We anticipate that these modifications will be particularly useful in studying thin films, superlattices or topological textures, where the intricate interplay of interactions requires a more precise treatment.
It is worth noting that an analytical treatment of dipole-dipole interactions is easy to incorporate whenever a reference structure is used to build the model and is inherently included for instance in effective Hamiltonian approaches or effective atomic potential approaches.
Strategies to incorporate the long-range part of the dipole-dipole interactions in machine learning potentials include charge equilibration schemes~\cite{Shaidu-24}, the on-the-fly computation of maximally localized Wannier function centers~\cite{Zhang-22,Gao-22}, or a recently proposed modification of the model without the need for additional first-principles calculations or retraining discussed in Ref.~\cite{Monacelli-24}.

\section*{Data availability}
All data are available in the main text. The second-principles model is open-source and available at Ref~\cite{dataverse_BTO_miao} along with a validation assessment.
\\

\acknowledgments
This work has been supported  by the European Union’s Horizon 2020 research and innovation program under Grant Agreement No. 964931 (TSAR) and by F.R.S.-FNRS Belgium under PDR grants T.0107.20 (PROMOSPAN) and T.0128.25 (TOPOTEX). M.Y. acknowledges financial support of the China Scholarship Council program (Grant No.202306240141). F.G.O. acknowledges financial support from MSCA-PF 101148906 funded by the European Union and the Fonds de la Recherche Scientifique (FNRS) through the grant FNRS-CR 1.B.227.25F. J. Z. acknowledges financial support from National Natural Science Foundation of China (NSFC, Grants 12274145), Guangdong Basic and Applied Basic Research Foundation, China (Grant 2023A1515010672), and Guangdong Provincial University Science and Technology Program (Grant 2023KTSCX029). The authors also acknowledge access to the Consortium des Équipements de Calcul Intensif (CÉCI), funded by the F.R.S.-FNRS under Grant No. 2.5020.11 and to the Tier-1 Lucia supercomputer of the Walloon Region, infrastructure funded by the Walloon Region under the grant agreement No. 1910247.  
%

\begin{thebibliography}{57}%
\makeatletter
\providecommand \@ifxundefined [1]{%
 \@ifx{#1\undefined}
}%
\providecommand \@ifnum [1]{%
 \ifnum #1\expandafter \@firstoftwo
 \else \expandafter \@secondoftwo
 \fi
}%
\providecommand \@ifx [1]{%
 \ifx #1\expandafter \@firstoftwo
 \else \expandafter \@secondoftwo
 \fi
}%
\providecommand \natexlab [1]{#1}%
\providecommand \enquote  [1]{``#1''}%
\providecommand \bibnamefont  [1]{#1}%
\providecommand \bibfnamefont [1]{#1}%
\providecommand \citenamefont [1]{#1}%
\providecommand \href@noop [0]{\@secondoftwo}%
\providecommand \href [0]{\begingroup \@sanitize@url \@href}%
\providecommand \@href[1]{\@@startlink{#1}\@@href}%
\providecommand \@@href[1]{\endgroup#1\@@endlink}%
\providecommand \@sanitize@url [0]{\catcode `\\12\catcode `\$12\catcode `\&12\catcode `\#12\catcode `\^12\catcode `\_12\catcode `\%12\relax}%
\providecommand \@@startlink[1]{}%
\providecommand \@@endlink[0]{}%
\providecommand \url  [0]{\begingroup\@sanitize@url \@url }%
\providecommand \@url [1]{\endgroup\@href {#1}{\urlprefix }}%
\providecommand \urlprefix  [0]{URL }%
\providecommand \Eprint [0]{\href }%
\providecommand \doibase [0]{https://doi.org/}%
\providecommand \selectlanguage [0]{\@gobble}%
\providecommand \bibinfo  [0]{\@secondoftwo}%
\providecommand \bibfield  [0]{\@secondoftwo}%
\providecommand \translation [1]{[#1]}%
\providecommand \BibitemOpen [0]{}%
\providecommand \bibitemStop [0]{}%
\providecommand \bibitemNoStop [0]{.\EOS\space}%
\providecommand \EOS [0]{\spacefactor3000\relax}%
\providecommand \BibitemShut  [1]{\csname bibitem#1\endcsname}%
\let\auto@bib@innerbib\@empty
\bibitem [{\citenamefont {Jones}\ and\ \citenamefont {Chapman}(1924)}]{Jones-24}%
  \BibitemOpen
  \bibfield  {author} {\bibinfo {author} {\bibfnamefont {J.~E.}\ \bibnamefont {Jones}}\ and\ \bibinfo {author} {\bibfnamefont {S.}~\bibnamefont {Chapman}},\ }\bibfield  {title} {\bibinfo {title} {On the determination of molecular fields. —{I}{I}. from the equation of state of a gas},\ }\href {https://doi.org/10.1098/rspa.1924.0082} {\bibfield  {journal} {\bibinfo  {journal} {Proceedings of the Royal Society of London. Series A, Containing Papers of a Mathematical and Physical Character}\ }\textbf {\bibinfo {volume} {106}},\ \bibinfo {pages} {463} (\bibinfo {year} {1924})}\BibitemShut {NoStop}%
\bibitem [{\citenamefont {Stillinger}\ and\ \citenamefont {Weber}(1985)}]{Weber-85}%
  \BibitemOpen
  \bibfield  {author} {\bibinfo {author} {\bibfnamefont {F.~H.}\ \bibnamefont {Stillinger}}\ and\ \bibinfo {author} {\bibfnamefont {T.~A.}\ \bibnamefont {Weber}},\ }\bibfield  {title} {\bibinfo {title} {Computer simulation of local order in condensed phases of silicon},\ }\href {https://doi.org/10.1103/PhysRevB.31.5262} {\bibfield  {journal} {\bibinfo  {journal} {Phys. Rev. B}\ }\textbf {\bibinfo {volume} {31}},\ \bibinfo {pages} {5262} (\bibinfo {year} {1985})}\BibitemShut {NoStop}%
\bibitem [{\citenamefont {Tersoff}(1988)}]{Tersoff-88}%
  \BibitemOpen
  \bibfield  {author} {\bibinfo {author} {\bibfnamefont {J.}~\bibnamefont {Tersoff}},\ }\bibfield  {title} {\bibinfo {title} {New empirical approach for the structure and energy of covalent systems},\ }\href {https://doi.org/10.1103/PhysRevB.37.6991} {\bibfield  {journal} {\bibinfo  {journal} {Phys. Rev. B}\ }\textbf {\bibinfo {volume} {37}},\ \bibinfo {pages} {6991} (\bibinfo {year} {1988})}\BibitemShut {NoStop}%
\bibitem [{\citenamefont {Pedone}\ \emph {et~al.}(2006)\citenamefont {Pedone}, \citenamefont {Malavasi}, \citenamefont {Menziani}, \citenamefont {Cormack},\ and\ \citenamefont {Segre}}]{Pedone2006}%
  \BibitemOpen
  \bibfield  {author} {\bibinfo {author} {\bibfnamefont {A.}~\bibnamefont {Pedone}}, \bibinfo {author} {\bibfnamefont {G.}~\bibnamefont {Malavasi}}, \bibinfo {author} {\bibfnamefont {M.~C.}\ \bibnamefont {Menziani}}, \bibinfo {author} {\bibfnamefont {A.~N.}\ \bibnamefont {Cormack}},\ and\ \bibinfo {author} {\bibfnamefont {U.}~\bibnamefont {Segre}},\ }\bibfield  {title} {\bibinfo {title} {A new self-consistent empirical interatomic potential model for oxides, silicates, and silica-based glasses},\ }\href {https://doi.org/10.1021/jp0611018} {\bibfield  {journal} {\bibinfo  {journal} {The Journal of Physical Chemistry B}\ }\textbf {\bibinfo {volume} {110}},\ \bibinfo {pages} {11780} (\bibinfo {year} {2006})}\BibitemShut {NoStop}%
\bibitem [{\citenamefont {Liang}\ \emph {et~al.}(2013)\citenamefont {Liang}, \citenamefont {Shin}, \citenamefont {Cheng}, \citenamefont {Yilmaz}, \citenamefont {Vishnu}, \citenamefont {Verners}, \citenamefont {Zou}, \citenamefont {Phillpot}, \citenamefont {Sinnott},\ and\ \citenamefont {van Duin}}]{Liang-13}%
  \BibitemOpen
  \bibfield  {author} {\bibinfo {author} {\bibfnamefont {T.}~\bibnamefont {Liang}}, \bibinfo {author} {\bibfnamefont {Y.~K.}\ \bibnamefont {Shin}}, \bibinfo {author} {\bibfnamefont {Y.-T.}\ \bibnamefont {Cheng}}, \bibinfo {author} {\bibfnamefont {D.~E.}\ \bibnamefont {Yilmaz}}, \bibinfo {author} {\bibfnamefont {K.~G.}\ \bibnamefont {Vishnu}}, \bibinfo {author} {\bibfnamefont {O.}~\bibnamefont {Verners}}, \bibinfo {author} {\bibfnamefont {C.}~\bibnamefont {Zou}}, \bibinfo {author} {\bibfnamefont {S.~R.}\ \bibnamefont {Phillpot}}, \bibinfo {author} {\bibfnamefont {S.~B.}\ \bibnamefont {Sinnott}},\ and\ \bibinfo {author} {\bibfnamefont {A.~C.}\ \bibnamefont {van Duin}},\ }\bibfield  {title} {\bibinfo {title} {Reactive potentials for advanced atomistic simulations},\ }\href {https://doi.org/https://doi.org/10.1146/annurev-matsci-071312-121610} {\bibfield  {journal} {\bibinfo  {journal} {Annual Review of Materials Research}\ }\textbf {\bibinfo {volume} {43}},\ \bibinfo {pages} {109} (\bibinfo {year}
  {2013})}\BibitemShut {NoStop}%
\bibitem [{\citenamefont {Harrison}\ \emph {et~al.}(2018)\citenamefont {Harrison}, \citenamefont {Schall}, \citenamefont {Maskey}, \citenamefont {Mikulski}, \citenamefont {Knippenberg},\ and\ \citenamefont {Morrow}}]{Harrison-18}%
  \BibitemOpen
  \bibfield  {author} {\bibinfo {author} {\bibfnamefont {J.~A.}\ \bibnamefont {Harrison}}, \bibinfo {author} {\bibfnamefont {J.~D.}\ \bibnamefont {Schall}}, \bibinfo {author} {\bibfnamefont {S.}~\bibnamefont {Maskey}}, \bibinfo {author} {\bibfnamefont {P.~T.}\ \bibnamefont {Mikulski}}, \bibinfo {author} {\bibfnamefont {M.~T.}\ \bibnamefont {Knippenberg}},\ and\ \bibinfo {author} {\bibfnamefont {B.~H.}\ \bibnamefont {Morrow}},\ }\bibfield  {title} {\bibinfo {title} {Review of force fields and intermolecular potentials used in atomistic computational materials research},\ }\href {https://doi.org/10.1063/1.5020808} {\bibfield  {journal} {\bibinfo  {journal} {Applied Physics Reviews}\ }\textbf {\bibinfo {volume} {5}},\ \bibinfo {pages} {031104} (\bibinfo {year} {2018})}\BibitemShut {NoStop}%
\bibitem [{\citenamefont {Kohn}\ and\ \citenamefont {Sham}(1965)}]{Kohn-65}%
  \BibitemOpen
  \bibfield  {author} {\bibinfo {author} {\bibfnamefont {W.}~\bibnamefont {Kohn}}\ and\ \bibinfo {author} {\bibfnamefont {L.~J.}\ \bibnamefont {Sham}},\ }\bibfield  {title} {\bibinfo {title} {Self-consistent equations including exchange and correlation effects},\ }\href {https://doi.org/10.1103/PhysRev.140.A1133} {\bibfield  {journal} {\bibinfo  {journal} {Phys. Rev.}\ }\textbf {\bibinfo {volume} {140}},\ \bibinfo {pages} {A1133} (\bibinfo {year} {1965})}\BibitemShut {NoStop}%
\bibitem [{\citenamefont {Hohenberg}\ and\ \citenamefont {Kohn}(1964)}]{Hohenberg-65}%
  \BibitemOpen
  \bibfield  {author} {\bibinfo {author} {\bibfnamefont {P.}~\bibnamefont {Hohenberg}}\ and\ \bibinfo {author} {\bibfnamefont {W.}~\bibnamefont {Kohn}},\ }\bibfield  {title} {\bibinfo {title} {Inhomogeneous electron gas},\ }\href {https://doi.org/10.1103/PhysRev.136.B864} {\bibfield  {journal} {\bibinfo  {journal} {Phys. Rev.}\ }\textbf {\bibinfo {volume} {136}},\ \bibinfo {pages} {B864} (\bibinfo {year} {1964})}\BibitemShut {NoStop}%
\bibitem [{\citenamefont {Perdew}\ \emph {et~al.}(1996)\citenamefont {Perdew}, \citenamefont {Burke},\ and\ \citenamefont {Ernzerhof}}]{Perdew-96}%
  \BibitemOpen
  \bibfield  {author} {\bibinfo {author} {\bibfnamefont {J.~P.}\ \bibnamefont {Perdew}}, \bibinfo {author} {\bibfnamefont {K.}~\bibnamefont {Burke}},\ and\ \bibinfo {author} {\bibfnamefont {M.}~\bibnamefont {Ernzerhof}},\ }\bibfield  {title} {\bibinfo {title} {Generalized gradient approximation made simple},\ }\href {https://doi.org/10.1103/PhysRevLett.77.3865} {\bibfield  {journal} {\bibinfo  {journal} {Phys. Rev. Lett.}\ }\textbf {\bibinfo {volume} {77}},\ \bibinfo {pages} {3865} (\bibinfo {year} {1996})}\BibitemShut {NoStop}%
\bibitem [{\citenamefont {Payne}\ \emph {et~al.}(1992)\citenamefont {Payne}, \citenamefont {Teter}, \citenamefont {Allan}, \citenamefont {Arias},\ and\ \citenamefont {Joannopoulos}}]{Payne-92}%
  \BibitemOpen
  \bibfield  {author} {\bibinfo {author} {\bibfnamefont {M.~C.}\ \bibnamefont {Payne}}, \bibinfo {author} {\bibfnamefont {M.~P.}\ \bibnamefont {Teter}}, \bibinfo {author} {\bibfnamefont {D.~C.}\ \bibnamefont {Allan}}, \bibinfo {author} {\bibfnamefont {T.~A.}\ \bibnamefont {Arias}},\ and\ \bibinfo {author} {\bibfnamefont {J.~D.}\ \bibnamefont {Joannopoulos}},\ }\bibfield  {title} {\bibinfo {title} {Iterative minimization techniques for ab initio total-energy calculations: molecular dynamics and conjugate gradients},\ }\href {https://doi.org/10.1103/RevModPhys.64.1045} {\bibfield  {journal} {\bibinfo  {journal} {Rev. Mod. Phys.}\ }\textbf {\bibinfo {volume} {64}},\ \bibinfo {pages} {1045} (\bibinfo {year} {1992})}\BibitemShut {NoStop}%
\bibitem [{\citenamefont {Ghosez}\ and\ \citenamefont {Junquera}(2022)}]{Ghosez-22}%
  \BibitemOpen
  \bibfield  {author} {\bibinfo {author} {\bibfnamefont {P.}~\bibnamefont {Ghosez}}\ and\ \bibinfo {author} {\bibfnamefont {J.}~\bibnamefont {Junquera}},\ }\bibfield  {title} {\bibinfo {title} {Modeling of ferroelectric oxide perovskites: From first to second principles},\ }\href {https://doi.org/https://doi.org/10.1146/annurev-conmatphys-040220-045528} {\bibfield  {journal} {\bibinfo  {journal} {Annual Review of Condensed Matter Physics}\ }\textbf {\bibinfo {volume} {13}},\ \bibinfo {pages} {325} (\bibinfo {year} {2022})}\BibitemShut {NoStop}%
\bibitem [{\citenamefont {Behler}\ and\ \citenamefont {Parrinello}(2007)}]{Behler-07}%
  \BibitemOpen
  \bibfield  {author} {\bibinfo {author} {\bibfnamefont {J.}~\bibnamefont {Behler}}\ and\ \bibinfo {author} {\bibfnamefont {M.}~\bibnamefont {Parrinello}},\ }\bibfield  {title} {\bibinfo {title} {Generalized neural-network representation of high-dimensional potential-energy surfaces},\ }\href {https://doi.org/10.1103/PhysRevLett.98.146401} {\bibfield  {journal} {\bibinfo  {journal} {Phys. Rev. Lett.}\ }\textbf {\bibinfo {volume} {98}},\ \bibinfo {pages} {146401} (\bibinfo {year} {2007})}\BibitemShut {NoStop}%
\bibitem [{\citenamefont {Chmiela}\ \emph {et~al.}(2017)\citenamefont {Chmiela}, \citenamefont {Tkatchenko}, \citenamefont {Sauceda}, \citenamefont {Poltavsky}, \citenamefont {Schütt},\ and\ \citenamefont {Müller}}]{Chmiela-17}%
  \BibitemOpen
  \bibfield  {author} {\bibinfo {author} {\bibfnamefont {S.}~\bibnamefont {Chmiela}}, \bibinfo {author} {\bibfnamefont {A.}~\bibnamefont {Tkatchenko}}, \bibinfo {author} {\bibfnamefont {H.~E.}\ \bibnamefont {Sauceda}}, \bibinfo {author} {\bibfnamefont {I.}~\bibnamefont {Poltavsky}}, \bibinfo {author} {\bibfnamefont {K.~T.}\ \bibnamefont {Schütt}},\ and\ \bibinfo {author} {\bibfnamefont {K.-R.}\ \bibnamefont {Müller}},\ }\bibfield  {title} {\bibinfo {title} {Machine learning of accurate energy-conserving molecular force fields},\ }\href {https://doi.org/10.1126/sciadv.1603015} {\bibfield  {journal} {\bibinfo  {journal} {Science Advances}\ }\textbf {\bibinfo {volume} {3}},\ \bibinfo {pages} {e1603015} (\bibinfo {year} {2017})}\BibitemShut {NoStop}%
\bibitem [{\citenamefont {Jinnouchi}\ \emph {et~al.}(2019)\citenamefont {Jinnouchi}, \citenamefont {Lahnsteiner}, \citenamefont {Karsai}, \citenamefont {Kresse},\ and\ \citenamefont {Bokdam}}]{Jinnouchi-19}%
  \BibitemOpen
  \bibfield  {author} {\bibinfo {author} {\bibfnamefont {R.}~\bibnamefont {Jinnouchi}}, \bibinfo {author} {\bibfnamefont {J.}~\bibnamefont {Lahnsteiner}}, \bibinfo {author} {\bibfnamefont {F.}~\bibnamefont {Karsai}}, \bibinfo {author} {\bibfnamefont {G.}~\bibnamefont {Kresse}},\ and\ \bibinfo {author} {\bibfnamefont {M.}~\bibnamefont {Bokdam}},\ }\bibfield  {title} {\bibinfo {title} {Phase transitions of hybrid perovskites simulated by machine-learning force fields trained on the fly with bayesian inference},\ }\href {https://doi.org/10.1103/PhysRevLett.122.225701} {\bibfield  {journal} {\bibinfo  {journal} {Phys. Rev. Lett.}\ }\textbf {\bibinfo {volume} {122}},\ \bibinfo {pages} {225701} (\bibinfo {year} {2019})}\BibitemShut {NoStop}%
\bibitem [{\citenamefont {Gorsse}\ \emph {et~al.}(2025)\citenamefont {Gorsse}, \citenamefont {Lin}, \citenamefont {Murakami}, \citenamefont {Rignanese},\ and\ \citenamefont {Yeh}}]{Grosse-25}%
  \BibitemOpen
  \bibfield  {author} {\bibinfo {author} {\bibfnamefont {S.}~\bibnamefont {Gorsse}}, \bibinfo {author} {\bibfnamefont {W.-C.}\ \bibnamefont {Lin}}, \bibinfo {author} {\bibfnamefont {H.}~\bibnamefont {Murakami}}, \bibinfo {author} {\bibfnamefont {G.-M.}\ \bibnamefont {Rignanese}},\ and\ \bibinfo {author} {\bibfnamefont {A.-C.}\ \bibnamefont {Yeh}},\ }\bibfield  {title} {\bibinfo {title} {Advancing refractory high entropy alloy development with ai-predictive models for high temperature oxidation resistance},\ }\href {https://doi.org/https://doi.org/10.1016/j.scriptamat.2024.116394} {\bibfield  {journal} {\bibinfo  {journal} {Scripta Materialia}\ }\textbf {\bibinfo {volume} {255}},\ \bibinfo {pages} {116394} (\bibinfo {year} {2025})}\BibitemShut {NoStop}%
\bibitem [{\citenamefont {Zhong}\ \emph {et~al.}(1994{\natexlab{a}})\citenamefont {Zhong}, \citenamefont {Vanderbilt},\ and\ \citenamefont {Rabe}}]{Zhong-94}%
  \BibitemOpen
  \bibfield  {author} {\bibinfo {author} {\bibfnamefont {W.}~\bibnamefont {Zhong}}, \bibinfo {author} {\bibfnamefont {D.}~\bibnamefont {Vanderbilt}},\ and\ \bibinfo {author} {\bibfnamefont {K.~M.}\ \bibnamefont {Rabe}},\ }\bibfield  {title} {\bibinfo {title} {Phase transitions in {B}a{T}i{O}$_{3}$ from first principles},\ }\href {https://doi.org/10.1103/PhysRevLett.73.1861} {\bibfield  {journal} {\bibinfo  {journal} {Phys. Rev. Lett.}\ }\textbf {\bibinfo {volume} {73}},\ \bibinfo {pages} {1861} (\bibinfo {year} {1994}{\natexlab{a}})}\BibitemShut {NoStop}%
\bibitem [{\citenamefont {Rabe}\ and\ \citenamefont {Waghmare}(1995)}]{Rabe-95}%
  \BibitemOpen
  \bibfield  {author} {\bibinfo {author} {\bibfnamefont {K.~M.}\ \bibnamefont {Rabe}}\ and\ \bibinfo {author} {\bibfnamefont {U.~V.}\ \bibnamefont {Waghmare}},\ }\bibfield  {title} {\bibinfo {title} {Localized basis for effective lattice hamiltonians: Lattice {W}annier functions},\ }\href {https://doi.org/10.1103/PhysRevB.52.13236} {\bibfield  {journal} {\bibinfo  {journal} {Phys. Rev. B}\ }\textbf {\bibinfo {volume} {52}},\ \bibinfo {pages} {13236} (\bibinfo {year} {1995})}\BibitemShut {NoStop}%
\bibitem [{\citenamefont {Wojdeł}\ \emph {et~al.}(2013)\citenamefont {Wojdeł}, \citenamefont {Hermet}, \citenamefont {Ljungberg}, \citenamefont {Ghosez},\ and\ \citenamefont {Íñiguez}}]{Wojdel-13}%
  \BibitemOpen
  \bibfield  {author} {\bibinfo {author} {\bibfnamefont {J.~C.}\ \bibnamefont {Wojdeł}}, \bibinfo {author} {\bibfnamefont {P.}~\bibnamefont {Hermet}}, \bibinfo {author} {\bibfnamefont {M.~P.}\ \bibnamefont {Ljungberg}}, \bibinfo {author} {\bibfnamefont {P.}~\bibnamefont {Ghosez}},\ and\ \bibinfo {author} {\bibfnamefont {J.}~\bibnamefont {Íñiguez}},\ }\bibfield  {title} {\bibinfo {title} {First-principles model potentials for lattice-dynamical studies: general methodology and example of application to ferroic perovskite oxides},\ }\href {https://doi.org/10.1088/0953-8984/25/30/305401} {\bibfield  {journal} {\bibinfo  {journal} {Journal of Physics: Condensed Matter}\ }\textbf {\bibinfo {volume} {25}},\ \bibinfo {pages} {305401} (\bibinfo {year} {2013})}\BibitemShut {NoStop}%
\bibitem [{\citenamefont {Escorihuela-Sayalero}\ \emph {et~al.}(2017)\citenamefont {Escorihuela-Sayalero}, \citenamefont {Wojde\l{}},\ and\ \citenamefont {\'I\~niguez}}]{Escorihuela-17}%
  \BibitemOpen
  \bibfield  {author} {\bibinfo {author} {\bibfnamefont {C.}~\bibnamefont {Escorihuela-Sayalero}}, \bibinfo {author} {\bibfnamefont {J.~C.}\ \bibnamefont {Wojde\l{}}},\ and\ \bibinfo {author} {\bibfnamefont {J.}~\bibnamefont {\'I\~niguez}},\ }\bibfield  {title} {\bibinfo {title} {Efficient systematic scheme to construct second-principles lattice dynamical models},\ }\href {https://doi.org/10.1103/PhysRevB.95.094115} {\bibfield  {journal} {\bibinfo  {journal} {Phys. Rev. B}\ }\textbf {\bibinfo {volume} {95}},\ \bibinfo {pages} {094115} (\bibinfo {year} {2017})}\BibitemShut {NoStop}%
\bibitem [{\citenamefont {Garc\'{\i}a-Fern\'andez}\ \emph {et~al.}(2016)\citenamefont {Garc\'{\i}a-Fern\'andez}, \citenamefont {Wojde\l{}}, \citenamefont {\'I\~niguez},\ and\ \citenamefont {Junquera}}]{García-Fernández-16}%
  \BibitemOpen
  \bibfield  {author} {\bibinfo {author} {\bibfnamefont {P.}~\bibnamefont {Garc\'{\i}a-Fern\'andez}}, \bibinfo {author} {\bibfnamefont {J.~C.}\ \bibnamefont {Wojde\l{}}}, \bibinfo {author} {\bibfnamefont {J.}~\bibnamefont {\'I\~niguez}},\ and\ \bibinfo {author} {\bibfnamefont {J.}~\bibnamefont {Junquera}},\ }\bibfield  {title} {\bibinfo {title} {Second-principles method for materials simulations including electron and lattice degrees of freedom},\ }\href {https://doi.org/10.1103/PhysRevB.93.195137} {\bibfield  {journal} {\bibinfo  {journal} {Phys. Rev. B}\ }\textbf {\bibinfo {volume} {93}},\ \bibinfo {pages} {195137} (\bibinfo {year} {2016})}\BibitemShut {NoStop}%
\bibitem [{\citenamefont {Gonze}\ \emph {et~al.}(2020)\citenamefont {Gonze}, \citenamefont {Amadon}, \citenamefont {Antonius}, \citenamefont {Arnardi}, \citenamefont {Baguet}, \citenamefont {Beuken}, \citenamefont {Bieder}, \citenamefont {Bottin}, \citenamefont {Bouchet}, \citenamefont {Bousquet} \emph {et~al.}}]{Gonze-20}%
  \BibitemOpen
  \bibfield  {author} {\bibinfo {author} {\bibfnamefont {X.}~\bibnamefont {Gonze}}, \bibinfo {author} {\bibfnamefont {B.}~\bibnamefont {Amadon}}, \bibinfo {author} {\bibfnamefont {G.}~\bibnamefont {Antonius}}, \bibinfo {author} {\bibfnamefont {F.}~\bibnamefont {Arnardi}}, \bibinfo {author} {\bibfnamefont {L.}~\bibnamefont {Baguet}}, \bibinfo {author} {\bibfnamefont {J.-M.}\ \bibnamefont {Beuken}}, \bibinfo {author} {\bibfnamefont {J.}~\bibnamefont {Bieder}}, \bibinfo {author} {\bibfnamefont {F.}~\bibnamefont {Bottin}}, \bibinfo {author} {\bibfnamefont {J.}~\bibnamefont {Bouchet}}, \bibinfo {author} {\bibfnamefont {E.}~\bibnamefont {Bousquet}}, \emph {et~al.},\ }\bibfield  {title} {\bibinfo {title} {The abinit project: Impact, environment and recent developments},\ }\href {https://doi.org/10.1016/j.cpc.2019.107042} {\bibfield  {journal} {\bibinfo  {journal} {Computer Physics Communications}\ }\textbf {\bibinfo {volume} {248}},\ \bibinfo {pages} {107042} (\bibinfo {year} {2020})}\BibitemShut {NoStop}%
\bibitem [{\citenamefont {Chen}(2002)}]{Long-Qing}%
  \BibitemOpen
  \bibfield  {author} {\bibinfo {author} {\bibfnamefont {L.-Q.}\ \bibnamefont {Chen}},\ }\bibfield  {title} {\bibinfo {title} {Phase-field models for microstructure evolution},\ }\href {https://doi.org/https://doi.org/10.1146/annurev.matsci.32.112001.132041} {\bibfield  {journal} {\bibinfo  {journal} {Annual Review of Materials Research}\ }\textbf {\bibinfo {volume} {32}},\ \bibinfo {pages} {113} (\bibinfo {year} {2002})}\BibitemShut {NoStop}%
\bibitem [{\citenamefont {Sepliarsky}\ \emph {et~al.}(2000)\citenamefont {Sepliarsky}, \citenamefont {Phillpot}, \citenamefont {Wolf}, \citenamefont {Stachiotti},\ and\ \citenamefont {Migoni}}]{Shell}%
  \BibitemOpen
  \bibfield  {author} {\bibinfo {author} {\bibfnamefont {M.}~\bibnamefont {Sepliarsky}}, \bibinfo {author} {\bibfnamefont {S.~R.}\ \bibnamefont {Phillpot}}, \bibinfo {author} {\bibfnamefont {D.}~\bibnamefont {Wolf}}, \bibinfo {author} {\bibfnamefont {M.~G.}\ \bibnamefont {Stachiotti}},\ and\ \bibinfo {author} {\bibfnamefont {R.~L.}\ \bibnamefont {Migoni}},\ }\bibfield  {title} {\bibinfo {title} {Atomic-level simulation of ferroelectricity in perovskite solid solutions},\ }\href {https://doi.org/10.1063/1.126843} {\bibfield  {journal} {\bibinfo  {journal} {Applied Physics Letters}\ }\textbf {\bibinfo {volume} {76}},\ \bibinfo {pages} {3986} (\bibinfo {year} {2000})}\BibitemShut {NoStop}%
\bibitem [{\citenamefont {Shin}\ \emph {et~al.}(2005)\citenamefont {Shin}, \citenamefont {Cooper}, \citenamefont {Grinberg},\ and\ \citenamefont {Rappe}}]{Bondv}%
  \BibitemOpen
  \bibfield  {author} {\bibinfo {author} {\bibfnamefont {Y.-H.}\ \bibnamefont {Shin}}, \bibinfo {author} {\bibfnamefont {V.~R.}\ \bibnamefont {Cooper}}, \bibinfo {author} {\bibfnamefont {I.}~\bibnamefont {Grinberg}},\ and\ \bibinfo {author} {\bibfnamefont {A.~M.}\ \bibnamefont {Rappe}},\ }\bibfield  {title} {\bibinfo {title} {Development of a bond-valence molecular-dynamics model for complex oxides},\ }\href {https://doi.org/10.1103/PhysRevB.71.054104} {\bibfield  {journal} {\bibinfo  {journal} {Phys. Rev. B}\ }\textbf {\bibinfo {volume} {71}},\ \bibinfo {pages} {054104} (\bibinfo {year} {2005})}\BibitemShut {NoStop}%
\bibitem [{\citenamefont {van Duin}\ \emph {et~al.}(2001)\citenamefont {van Duin}, \citenamefont {Dasgupta}, \citenamefont {Lorant},\ and\ \citenamefont {Goddard}}]{vanDuin-01}%
  \BibitemOpen
  \bibfield  {author} {\bibinfo {author} {\bibfnamefont {A.~C.~T.}\ \bibnamefont {van Duin}}, \bibinfo {author} {\bibfnamefont {S.}~\bibnamefont {Dasgupta}}, \bibinfo {author} {\bibfnamefont {F.}~\bibnamefont {Lorant}},\ and\ \bibinfo {author} {\bibfnamefont {W.~A.}\ \bibnamefont {Goddard}},\ }\bibfield  {title} {\bibinfo {title} {Reax{F}{F}:{\thinspace} a reactive force field for hydrocarbons},\ }\href {https://doi.org/10.1021/jp004368u} {\bibfield  {journal} {\bibinfo  {journal} {The Journal of Physical Chemistry A}\ }\textbf {\bibinfo {volume} {105}},\ \bibinfo {pages} {9396} (\bibinfo {year} {2001})}\BibitemShut {NoStop}%
\bibitem [{\citenamefont {Akbarian}\ \emph {et~al.}(2019)\citenamefont {Akbarian}, \citenamefont {Yilmaz}, \citenamefont {Cao}, \citenamefont {Ganesh}, \citenamefont {Dabo}, \citenamefont {Munro}, \citenamefont {Van~Ginhoven},\ and\ \citenamefont {van Duin}}]{Ganesh-19}%
  \BibitemOpen
  \bibfield  {author} {\bibinfo {author} {\bibfnamefont {D.}~\bibnamefont {Akbarian}}, \bibinfo {author} {\bibfnamefont {D.~E.}\ \bibnamefont {Yilmaz}}, \bibinfo {author} {\bibfnamefont {Y.}~\bibnamefont {Cao}}, \bibinfo {author} {\bibfnamefont {P.}~\bibnamefont {Ganesh}}, \bibinfo {author} {\bibfnamefont {I.}~\bibnamefont {Dabo}}, \bibinfo {author} {\bibfnamefont {J.}~\bibnamefont {Munro}}, \bibinfo {author} {\bibfnamefont {R.}~\bibnamefont {Van~Ginhoven}},\ and\ \bibinfo {author} {\bibfnamefont {A.~C.~T.}\ \bibnamefont {van Duin}},\ }\bibfield  {title} {\bibinfo {title} {Understanding the influence of defects and surface chemistry on ferroelectric switching: a reax{F}{F} investigation of {B}a{T}i{O}$_3$},\ }\href {https://doi.org/10.1039/C9CP02955A} {\bibfield  {journal} {\bibinfo  {journal} {Phys. Chem. Chem. Phys.}\ }\textbf {\bibinfo {volume} {21}},\ \bibinfo {pages} {18240} (\bibinfo {year} {2019})}\BibitemShut {NoStop}%
\bibitem [{\citenamefont {Kelley}\ \emph {et~al.}(2022)\citenamefont {Kelley}, \citenamefont {Morozovska}, \citenamefont {Eliseev}, \citenamefont {Sharma}, \citenamefont {Yilmaz}, \citenamefont {van Duin}, \citenamefont {Ganesh}, \citenamefont {Borisevich}, \citenamefont {Jesse}, \citenamefont {Maksymovych}, \citenamefont {Balke}, \citenamefont {Kalinin},\ and\ \citenamefont {Vasudevan}}]{Ganesh-22}%
  \BibitemOpen
  \bibfield  {author} {\bibinfo {author} {\bibfnamefont {K.~P.}\ \bibnamefont {Kelley}}, \bibinfo {author} {\bibfnamefont {A.~N.}\ \bibnamefont {Morozovska}}, \bibinfo {author} {\bibfnamefont {E.~A.}\ \bibnamefont {Eliseev}}, \bibinfo {author} {\bibfnamefont {V.}~\bibnamefont {Sharma}}, \bibinfo {author} {\bibfnamefont {D.~E.}\ \bibnamefont {Yilmaz}}, \bibinfo {author} {\bibfnamefont {A.~C.~T.}\ \bibnamefont {van Duin}}, \bibinfo {author} {\bibfnamefont {P.}~\bibnamefont {Ganesh}}, \bibinfo {author} {\bibfnamefont {A.}~\bibnamefont {Borisevich}}, \bibinfo {author} {\bibfnamefont {S.}~\bibnamefont {Jesse}}, \bibinfo {author} {\bibfnamefont {P.}~\bibnamefont {Maksymovych}}, \bibinfo {author} {\bibfnamefont {N.}~\bibnamefont {Balke}}, \bibinfo {author} {\bibfnamefont {S.~V.}\ \bibnamefont {Kalinin}},\ and\ \bibinfo {author} {\bibfnamefont {R.~K.}\ \bibnamefont {Vasudevan}},\ }\bibfield  {title} {\bibinfo {title} {Oxygen vacancy injection as a pathway to enhancing electromechanical response in ferroelectrics},\
  }\href {https://doi.org/https://doi.org/10.1002/adma.202106426} {\bibfield  {journal} {\bibinfo  {journal} {Advanced Materials}\ }\textbf {\bibinfo {volume} {34}},\ \bibinfo {pages} {2106426} (\bibinfo {year} {2022})}\BibitemShut {NoStop}%
\bibitem [{\citenamefont {Dhakane}\ \emph {et~al.}(2023)\citenamefont {Dhakane}, \citenamefont {Xie}, \citenamefont {Yilmaz}, \citenamefont {van Duin}, \citenamefont {Sumpter},\ and\ \citenamefont {Ganesh}}]{Ganesh-23}%
  \BibitemOpen
  \bibfield  {author} {\bibinfo {author} {\bibfnamefont {A.}~\bibnamefont {Dhakane}}, \bibinfo {author} {\bibfnamefont {T.}~\bibnamefont {Xie}}, \bibinfo {author} {\bibfnamefont {D.~E.}\ \bibnamefont {Yilmaz}}, \bibinfo {author} {\bibfnamefont {A.~C.}\ \bibnamefont {van Duin}}, \bibinfo {author} {\bibfnamefont {B.~G.}\ \bibnamefont {Sumpter}},\ and\ \bibinfo {author} {\bibfnamefont {P.}~\bibnamefont {Ganesh}},\ }\bibfield  {title} {\bibinfo {title} {A graph dynamical neural network approach for decoding dynamical states in ferroelectrics.},\ }\href {https://doi.org/https://doi.org/10.1016/j.cartre.2023.100264} {\bibfield  {journal} {\bibinfo  {journal} {Carbon Trends}\ }\textbf {\bibinfo {volume} {11}},\ \bibinfo {pages} {100264} (\bibinfo {year} {2023})}\BibitemShut {NoStop}%
\bibitem [{\citenamefont {Wu}\ \emph {et~al.}(2023)\citenamefont {Wu}, \citenamefont {Yang}, \citenamefont {Liu}, \citenamefont {Zhang}, \citenamefont {Yang}, \citenamefont {Zhang}, \citenamefont {Zhang},\ and\ \citenamefont {Liu}}]{Universal}%
  \BibitemOpen
  \bibfield  {author} {\bibinfo {author} {\bibfnamefont {J.}~\bibnamefont {Wu}}, \bibinfo {author} {\bibfnamefont {J.}~\bibnamefont {Yang}}, \bibinfo {author} {\bibfnamefont {Y.-J.}\ \bibnamefont {Liu}}, \bibinfo {author} {\bibfnamefont {D.}~\bibnamefont {Zhang}}, \bibinfo {author} {\bibfnamefont {Y.}~\bibnamefont {Yang}}, \bibinfo {author} {\bibfnamefont {Y.}~\bibnamefont {Zhang}}, \bibinfo {author} {\bibfnamefont {L.}~\bibnamefont {Zhang}},\ and\ \bibinfo {author} {\bibfnamefont {S.}~\bibnamefont {Liu}},\ }\bibfield  {title} {\bibinfo {title} {Universal interatomic potential for perovskite oxides},\ }\href {https://doi.org/10.1103/PhysRevB.108.L180104} {\bibfield  {journal} {\bibinfo  {journal} {Phys. Rev. B}\ }\textbf {\bibinfo {volume} {108}},\ \bibinfo {pages} {L180104} (\bibinfo {year} {2023})}\BibitemShut {NoStop}%
\bibitem [{\citenamefont {Zhang}\ \emph {et~al.}(2018)\citenamefont {Zhang}, \citenamefont {Han}, \citenamefont {Wang}, \citenamefont {Car},\ and\ \citenamefont {E}}]{Zhang-18}%
  \BibitemOpen
  \bibfield  {author} {\bibinfo {author} {\bibfnamefont {L.}~\bibnamefont {Zhang}}, \bibinfo {author} {\bibfnamefont {J.}~\bibnamefont {Han}}, \bibinfo {author} {\bibfnamefont {H.}~\bibnamefont {Wang}}, \bibinfo {author} {\bibfnamefont {R.}~\bibnamefont {Car}},\ and\ \bibinfo {author} {\bibfnamefont {W.}~\bibnamefont {E}},\ }\bibfield  {title} {\bibinfo {title} {Deep potential molecular dynamics: A scalable model with the accuracy of quantum mechanics},\ }\href {https://doi.org/10.1103/PhysRevLett.120.143001} {\bibfield  {journal} {\bibinfo  {journal} {Phys. Rev. Lett.}\ }\textbf {\bibinfo {volume} {120}},\ \bibinfo {pages} {143001} (\bibinfo {year} {2018})}\BibitemShut {NoStop}%
\bibitem [{\citenamefont {Botu}\ and\ \citenamefont {Ramprasad}(2015)}]{Botu-15}%
  \BibitemOpen
  \bibfield  {author} {\bibinfo {author} {\bibfnamefont {V.}~\bibnamefont {Botu}}\ and\ \bibinfo {author} {\bibfnamefont {R.}~\bibnamefont {Ramprasad}},\ }\bibfield  {title} {\bibinfo {title} {Learning scheme to predict atomic forces and accelerate materials simulations},\ }\href {https://doi.org/10.1103/PhysRevB.92.094306} {\bibfield  {journal} {\bibinfo  {journal} {Phys. Rev. B}\ }\textbf {\bibinfo {volume} {92}},\ \bibinfo {pages} {094306} (\bibinfo {year} {2015})}\BibitemShut {NoStop}%
\bibitem [{\citenamefont {Klawohn}\ \emph {et~al.}(2023)\citenamefont {Klawohn}, \citenamefont {Darby}, \citenamefont {Kermode}, \citenamefont {Csányi}, \citenamefont {Caro},\ and\ \citenamefont {Bartók}}]{Klawohn-23}%
  \BibitemOpen
  \bibfield  {author} {\bibinfo {author} {\bibfnamefont {S.}~\bibnamefont {Klawohn}}, \bibinfo {author} {\bibfnamefont {J.~P.}\ \bibnamefont {Darby}}, \bibinfo {author} {\bibfnamefont {J.~R.}\ \bibnamefont {Kermode}}, \bibinfo {author} {\bibfnamefont {G.}~\bibnamefont {Csányi}}, \bibinfo {author} {\bibfnamefont {M.~A.}\ \bibnamefont {Caro}},\ and\ \bibinfo {author} {\bibfnamefont {A.~P.}\ \bibnamefont {Bartók}},\ }\bibfield  {title} {\bibinfo {title} {Gaussian approximation potentials: Theory, software implementation and application examples},\ }\href {https://doi.org/10.1063/5.0160898} {\bibfield  {journal} {\bibinfo  {journal} {The Journal of Chemical Physics}\ }\textbf {\bibinfo {volume} {159}},\ \bibinfo {pages} {174108} (\bibinfo {year} {2023})}\BibitemShut {NoStop}%
\bibitem [{\citenamefont {Singraber}\ \emph {et~al.}(2019)\citenamefont {Singraber}, \citenamefont {Behler},\ and\ \citenamefont {Dellago}}]{Singraber-19}%
  \BibitemOpen
  \bibfield  {author} {\bibinfo {author} {\bibfnamefont {A.}~\bibnamefont {Singraber}}, \bibinfo {author} {\bibfnamefont {J.}~\bibnamefont {Behler}},\ and\ \bibinfo {author} {\bibfnamefont {C.}~\bibnamefont {Dellago}},\ }\bibfield  {title} {\bibinfo {title} {Library-based lammps implementation of high-dimensional neural network potentials},\ }\href {https://doi.org/10.1021/acs.jctc.8b00770} {\bibfield  {journal} {\bibinfo  {journal} {Journal of Chemical Theory and Computation}\ }\textbf {\bibinfo {volume} {15}},\ \bibinfo {pages} {1827} (\bibinfo {year} {2019})}\BibitemShut {NoStop}%
\bibitem [{\citenamefont {Shaidu}\ \emph {et~al.}(2024)\citenamefont {Shaidu}, \citenamefont {Pellegrini}, \citenamefont {K{\"u}{\c{c}}{\"u}kbenli}, \citenamefont {Lot},\ and\ \citenamefont {de~Gironcoli}}]{Shaidu-24}%
  \BibitemOpen
  \bibfield  {author} {\bibinfo {author} {\bibfnamefont {Y.}~\bibnamefont {Shaidu}}, \bibinfo {author} {\bibfnamefont {F.}~\bibnamefont {Pellegrini}}, \bibinfo {author} {\bibfnamefont {E.}~\bibnamefont {K{\"u}{\c{c}}{\"u}kbenli}}, \bibinfo {author} {\bibfnamefont {R.}~\bibnamefont {Lot}},\ and\ \bibinfo {author} {\bibfnamefont {S.}~\bibnamefont {de~Gironcoli}},\ }\bibfield  {title} {\bibinfo {title} {Incorporating long-range electrostatics in neural network potentials via variational charge equilibration from shortsighted ingredients},\ }\href {https://doi.org/10.1038/s41524-024-01225-6} {\bibfield  {journal} {\bibinfo  {journal} {npj Computational Materials}\ }\textbf {\bibinfo {volume} {10}},\ \bibinfo {pages} {47} (\bibinfo {year} {2024})}\BibitemShut {NoStop}%
\bibitem [{\citenamefont {Zhang}\ \emph {et~al.}(2022)\citenamefont {Zhang}, \citenamefont {Wang}, \citenamefont {Muniz}, \citenamefont {Panagiotopoulos}, \citenamefont {Car},\ and\ \citenamefont {E}}]{Zhang-22}%
  \BibitemOpen
  \bibfield  {author} {\bibinfo {author} {\bibfnamefont {L.}~\bibnamefont {Zhang}}, \bibinfo {author} {\bibfnamefont {H.}~\bibnamefont {Wang}}, \bibinfo {author} {\bibfnamefont {M.~C.}\ \bibnamefont {Muniz}}, \bibinfo {author} {\bibfnamefont {A.~Z.}\ \bibnamefont {Panagiotopoulos}}, \bibinfo {author} {\bibfnamefont {R.}~\bibnamefont {Car}},\ and\ \bibinfo {author} {\bibfnamefont {W.}~\bibnamefont {E}},\ }\bibfield  {title} {\bibinfo {title} {A deep potential model with long-range electrostatic interactions},\ }\href {https://doi.org/10.1063/5.0083669} {\bibfield  {journal} {\bibinfo  {journal} {J. Chem. Phys.}\ }\textbf {\bibinfo {volume} {156}},\ \bibinfo {pages} {124107} (\bibinfo {year} {2022})}\BibitemShut {NoStop}%
\bibitem [{\citenamefont {Gao}\ and\ \citenamefont {Remsing}(2022)}]{Gao-22}%
  \BibitemOpen
  \bibfield  {author} {\bibinfo {author} {\bibfnamefont {A.}~\bibnamefont {Gao}}\ and\ \bibinfo {author} {\bibfnamefont {R.~C.}\ \bibnamefont {Remsing}},\ }\bibfield  {title} {\bibinfo {title} {Self-consistent determination of long-range electrostatics in neural network potentials},\ }\href {https://doi.org/10.1038/s41467-022-29243-2} {\bibfield  {journal} {\bibinfo  {journal} {Nature Communications}\ }\textbf {\bibinfo {volume} {13}},\ \bibinfo {pages} {1572} (\bibinfo {year} {2022})}\BibitemShut {NoStop}%
\bibitem [{\citenamefont {Monacelli}\ and\ \citenamefont {Marzari}(2024)}]{Monacelli-24}%
  \BibitemOpen
  \bibfield  {author} {\bibinfo {author} {\bibfnamefont {L.}~\bibnamefont {Monacelli}}\ and\ \bibinfo {author} {\bibfnamefont {N.}~\bibnamefont {Marzari}},\ }\href {https://arxiv.org/abs/2412.01642} {\bibinfo {title} {Electrostatic interactions in atomistic and machine-learned potentials for polar materials}} (\bibinfo {year} {2024}),\ \Eprint {https://arxiv.org/abs/2412.01642} {arXiv:2412.01642 [cond-mat.mtrl-sci]} \BibitemShut {NoStop}%
\bibitem [{\citenamefont {Cochran}(1960)}]{Cochran-60}%
  \BibitemOpen
  \bibfield  {author} {\bibinfo {author} {\bibfnamefont {W.}~\bibnamefont {Cochran}},\ }\bibfield  {title} {\bibinfo {title} {Crystal stability and the theory of ferroelectricity},\ }\href {https://doi.org/10.1080/00018736000101229} {\bibfield  {journal} {\bibinfo  {journal} {Advances in Physics}\ }\textbf {\bibinfo {volume} {9}},\ \bibinfo {pages} {387} (\bibinfo {year} {1960})}\BibitemShut {NoStop}%
\bibitem [{\citenamefont {Ghosez}\ \emph {et~al.}(1996)\citenamefont {Ghosez}, \citenamefont {Gonze},\ and\ \citenamefont {Michenaud}}]{Ghosez-96}%
  \BibitemOpen
  \bibfield  {author} {\bibinfo {author} {\bibfnamefont {P.}~\bibnamefont {Ghosez}}, \bibinfo {author} {\bibfnamefont {X.}~\bibnamefont {Gonze}},\ and\ \bibinfo {author} {\bibfnamefont {J.-P.}\ \bibnamefont {Michenaud}},\ }\bibfield  {title} {\bibinfo {title} {Coulomb interaction and ferroelectric instability of {B}a{T}i{O}$_{3}$},\ }\href {https://doi.org/10.1209/epl/i1996-00404-8} {\bibfield  {journal} {\bibinfo  {journal} {Europhysics Letters}\ }\textbf {\bibinfo {volume} {33}},\ \bibinfo {pages} {713} (\bibinfo {year} {1996})}\BibitemShut {NoStop}%
\bibitem [{\citenamefont {Gigli}\ \emph {et~al.}(2022)\citenamefont {Gigli}, \citenamefont {Veit}, \citenamefont {Kotiuga}, \citenamefont {Pizzi}, \citenamefont {Marzari},\ and\ \citenamefont {Ceriotti}}]{Gigli-22}%
  \BibitemOpen
  \bibfield  {author} {\bibinfo {author} {\bibfnamefont {L.}~\bibnamefont {Gigli}}, \bibinfo {author} {\bibfnamefont {M.}~\bibnamefont {Veit}}, \bibinfo {author} {\bibfnamefont {M.}~\bibnamefont {Kotiuga}}, \bibinfo {author} {\bibfnamefont {G.}~\bibnamefont {Pizzi}}, \bibinfo {author} {\bibfnamefont {N.}~\bibnamefont {Marzari}},\ and\ \bibinfo {author} {\bibfnamefont {M.}~\bibnamefont {Ceriotti}},\ }\bibfield  {title} {\bibinfo {title} {Thermodynamics and dielectric response of {B}a{T}i{O}$_{3}$ by data-driven modeling},\ }\href {https://doi.org/10.1038/s41524-022-00845-0} {\bibfield  {journal} {\bibinfo  {journal} {npj Computational Materials}\ }\textbf {\bibinfo {volume} {8}},\ \bibinfo {pages} {209} (\bibinfo {year} {2022})}\BibitemShut {NoStop}%
\bibitem [{\citenamefont {G\'omez-Ortiz}\ \emph {et~al.}(2025)\citenamefont {G\'omez-Ortiz}, \citenamefont {Bastogne}, \citenamefont {Anand}, \citenamefont {Yu}, \citenamefont {He},\ and\ \citenamefont {Ghosez}}]{Gomez-25}%
  \BibitemOpen
  \bibfield  {author} {\bibinfo {author} {\bibfnamefont {F.}~\bibnamefont {G\'omez-Ortiz}}, \bibinfo {author} {\bibfnamefont {L.}~\bibnamefont {Bastogne}}, \bibinfo {author} {\bibfnamefont {S.}~\bibnamefont {Anand}}, \bibinfo {author} {\bibfnamefont {M.}~\bibnamefont {Yu}}, \bibinfo {author} {\bibfnamefont {X.}~\bibnamefont {He}},\ and\ \bibinfo {author} {\bibfnamefont {P.}~\bibnamefont {Ghosez}},\ }\bibfield  {title} {\bibinfo {title} {Switchable skyrmion--antiskyrmion tubes in rhombohedral {B}a{T}i{O}$_{3}$ and related materials},\ }\href {https://doi.org/10.1103/PhysRevB.111.L180104} {\bibfield  {journal} {\bibinfo  {journal} {Phys. Rev. B}\ }\textbf {\bibinfo {volume} {111}},\ \bibinfo {pages} {L180104} (\bibinfo {year} {2025})}\BibitemShut {NoStop}%
\bibitem [{\citenamefont {Gonze}\ and\ \citenamefont {Lee}(1997)}]{Gonze-97}%
  \BibitemOpen
  \bibfield  {author} {\bibinfo {author} {\bibfnamefont {X.}~\bibnamefont {Gonze}}\ and\ \bibinfo {author} {\bibfnamefont {C.}~\bibnamefont {Lee}},\ }\bibfield  {title} {\bibinfo {title} {Dynamical matrices, born effective charges, dielectric permittivity tensors, and interatomic force constants from density-functional perturbation theory},\ }\href {https://doi.org/10.1103/PhysRevB.55.10355} {\bibfield  {journal} {\bibinfo  {journal} {Phys. Rev. B}\ }\textbf {\bibinfo {volume} {55}},\ \bibinfo {pages} {10355} (\bibinfo {year} {1997})}\BibitemShut {NoStop}%
\bibitem [{\citenamefont {Fletcher}(2000)}]{fletcher2000practical}%
  \BibitemOpen
  \bibfield  {author} {\bibinfo {author} {\bibfnamefont {R.}~\bibnamefont {Fletcher}},\ }\href@noop {} {\emph {\bibinfo {title} {Practical methods of optimization}}}\ (\bibinfo  {publisher} {John Wiley \& Sons},\ \bibinfo {year} {2000})\BibitemShut {NoStop}%
\bibitem [{\citenamefont {Duane}\ \emph {et~al.}(1987)\citenamefont {Duane}, \citenamefont {Kennedy}, \citenamefont {Pendleton},\ and\ \citenamefont {Roweth}}]{duane1987hybrid}%
  \BibitemOpen
  \bibfield  {author} {\bibinfo {author} {\bibfnamefont {S.}~\bibnamefont {Duane}}, \bibinfo {author} {\bibfnamefont {A.~D.}\ \bibnamefont {Kennedy}}, \bibinfo {author} {\bibfnamefont {B.~J.}\ \bibnamefont {Pendleton}},\ and\ \bibinfo {author} {\bibfnamefont {D.}~\bibnamefont {Roweth}},\ }\bibfield  {title} {\bibinfo {title} {Hybrid monte carlo},\ }\href {https://doi.org/10.1016/0370-2693(87)91197-X} {\bibfield  {journal} {\bibinfo  {journal} {Physics letters B}\ }\textbf {\bibinfo {volume} {195}},\ \bibinfo {pages} {216} (\bibinfo {year} {1987})}\BibitemShut {NoStop}%
\bibitem [{\citenamefont {Betancourt}(2017)}]{betancourt2017conceptual}%
  \BibitemOpen
  \bibfield  {author} {\bibinfo {author} {\bibfnamefont {M.}~\bibnamefont {Betancourt}},\ }\bibfield  {title} {\bibinfo {title} {A conceptual introduction to {H}amiltonian {M}onte {C}arlo},\ }\bibfield  {journal} {\bibinfo  {journal} {arXiv preprint arXiv:1701.02434}\ }\href {https://doi.org/10.48550/arXiv.1701.02434} {10.48550/arXiv.1701.02434} (\bibinfo {year} {2017})\BibitemShut {NoStop}%
\bibitem [{\citenamefont {Ph.~Ghosez}\ and\ \citenamefont {Michenaud}(1998)}]{Ghosez-98}%
  \BibitemOpen
  \bibfield  {author} {\bibinfo {author} {\bibfnamefont {X.~G.}\ \bibnamefont {Ph.~Ghosez}}\ and\ \bibinfo {author} {\bibfnamefont {J.~P.}\ \bibnamefont {Michenaud}},\ }\bibfield  {title} {\bibinfo {title} {Ab initio phonon dispersion curves and interatomic force constants of barium titanate},\ }\href {https://doi.org/10.1080/00150199808009159} {\bibfield  {journal} {\bibinfo  {journal} {Ferroelectrics}\ }\textbf {\bibinfo {volume} {206}},\ \bibinfo {pages} {205} (\bibinfo {year} {1998})}\BibitemShut {NoStop}%
\bibitem [{\citenamefont {Chen}\ \emph {et~al.}(2024)\citenamefont {Chen}, \citenamefont {Zhao},\ and\ \citenamefont {Ghosez}}]{Chen-24}%
  \BibitemOpen
  \bibfield  {author} {\bibinfo {author} {\bibfnamefont {X.-L.}\ \bibnamefont {Chen}}, \bibinfo {author} {\bibfnamefont {J.-Z.}\ \bibnamefont {Zhao}},\ and\ \bibinfo {author} {\bibfnamefont {P.}~\bibnamefont {Ghosez}},\ }\bibfield  {title} {\bibinfo {title} {Lattice dynamical properties and interatomic force constants of transition metal oxide perovskite superlattices},\ }\href {https://doi.org/10.1103/PhysRevB.110.245302} {\bibfield  {journal} {\bibinfo  {journal} {Phys. Rev. B}\ }\textbf {\bibinfo {volume} {110}},\ \bibinfo {pages} {245302} (\bibinfo {year} {2024})}\BibitemShut {NoStop}%
\bibitem [{\citenamefont {Zhong}\ \emph {et~al.}(1994{\natexlab{b}})\citenamefont {Zhong}, \citenamefont {King-Smith},\ and\ \citenamefont {Vanderbilt}}]{Zhong-94.2}%
  \BibitemOpen
  \bibfield  {author} {\bibinfo {author} {\bibfnamefont {W.}~\bibnamefont {Zhong}}, \bibinfo {author} {\bibfnamefont {R.~D.}\ \bibnamefont {King-Smith}},\ and\ \bibinfo {author} {\bibfnamefont {D.}~\bibnamefont {Vanderbilt}},\ }\bibfield  {title} {\bibinfo {title} {Giant {L}{O}-{T}{O} splittings in perovskite ferroelectrics},\ }\href {https://doi.org/10.1103/PhysRevLett.72.3618} {\bibfield  {journal} {\bibinfo  {journal} {Phys. Rev. Lett.}\ }\textbf {\bibinfo {volume} {72}},\ \bibinfo {pages} {3618} (\bibinfo {year} {1994}{\natexlab{b}})}\BibitemShut {NoStop}%
\bibitem [{\citenamefont {Ph.~Ghosez}\ and\ \citenamefont {Michenaud}(1997)}]{Ghosez-97}%
  \BibitemOpen
  \bibfield  {author} {\bibinfo {author} {\bibfnamefont {X.~G.}\ \bibnamefont {Ph.~Ghosez}}\ and\ \bibinfo {author} {\bibfnamefont {J.-P.}\ \bibnamefont {Michenaud}},\ }\bibfield  {title} {\bibinfo {title} {Lattice dynamics and ferroelectric instability of barium titanate},\ }\href {https://doi.org/10.1080/00150199708016081} {\bibfield  {journal} {\bibinfo  {journal} {Ferroelectrics}\ }\textbf {\bibinfo {volume} {194}},\ \bibinfo {pages} {39} (\bibinfo {year} {1997})}\BibitemShut {NoStop}%
\bibitem [{\citenamefont {Ghosez}\ \emph {et~al.}(1999)\citenamefont {Ghosez}, \citenamefont {Cockayne}, \citenamefont {Waghmare},\ and\ \citenamefont {Rabe}}]{Ghosez-99}%
  \BibitemOpen
  \bibfield  {author} {\bibinfo {author} {\bibfnamefont {P.}~\bibnamefont {Ghosez}}, \bibinfo {author} {\bibfnamefont {E.}~\bibnamefont {Cockayne}}, \bibinfo {author} {\bibfnamefont {U.~V.}\ \bibnamefont {Waghmare}},\ and\ \bibinfo {author} {\bibfnamefont {K.~M.}\ \bibnamefont {Rabe}},\ }\bibfield  {title} {\bibinfo {title} {Lattice dynamics of {B}a{T}i{O}$_{3}$, {P}b{T}i{O}$_{3}$, and {P}b{Z}r{O}$_{3}$: A comparative first-principles study},\ }\href {https://doi.org/10.1103/PhysRevB.60.836} {\bibfield  {journal} {\bibinfo  {journal} {Phys. Rev. B}\ }\textbf {\bibinfo {volume} {60}},\ \bibinfo {pages} {836} (\bibinfo {year} {1999})}\BibitemShut {NoStop}%
\bibitem [{\citenamefont {Gigli}\ \emph {et~al.}(2024)\citenamefont {Gigli}, \citenamefont {Goscinski}, \citenamefont {Ceriotti},\ and\ \citenamefont {Tribello}}]{Gigli-24}%
  \BibitemOpen
  \bibfield  {author} {\bibinfo {author} {\bibfnamefont {L.}~\bibnamefont {Gigli}}, \bibinfo {author} {\bibfnamefont {A.}~\bibnamefont {Goscinski}}, \bibinfo {author} {\bibfnamefont {M.}~\bibnamefont {Ceriotti}},\ and\ \bibinfo {author} {\bibfnamefont {G.~A.}\ \bibnamefont {Tribello}},\ }\bibfield  {title} {\bibinfo {title} {Modeling the ferroelectric phase transition in barium titanate with dft accuracy and converged sampling},\ }\href {https://doi.org/10.1103/PhysRevB.110.024101} {\bibfield  {journal} {\bibinfo  {journal} {Phys. Rev. B}\ }\textbf {\bibinfo {volume} {110}},\ \bibinfo {pages} {024101} (\bibinfo {year} {2024})}\BibitemShut {NoStop}%
\bibitem [{\citenamefont {Zhang}\ and\ \citenamefont {Chen}(2025)}]{Zhang2025}%
  \BibitemOpen
  \bibfield  {author} {\bibinfo {author} {\bibfnamefont {C.}~\bibnamefont {Zhang}}\ and\ \bibinfo {author} {\bibfnamefont {X.}~\bibnamefont {Chen}},\ }\bibfield  {title} {\bibinfo {title} {Ferro{A}{I}: a deep learning model for predicting phase diagrams of ferroelectric materials},\ }\href {https://doi.org/10.1038/s41524-025-01778-0} {\bibfield  {journal} {\bibinfo  {journal} {npj Computational Materials}\ }\textbf {\bibinfo {volume} {11}},\ \bibinfo {pages} {282} (\bibinfo {year} {2025})}\BibitemShut {NoStop}%
\bibitem [{\citenamefont {Thong}\ \emph {et~al.}(2023)\citenamefont {Thong}, \citenamefont {Wang}, \citenamefont {Han}, \citenamefont {Zhang}, \citenamefont {Li}, \citenamefont {Wang},\ and\ \citenamefont {Xu}}]{Thong-23}%
  \BibitemOpen
  \bibfield  {author} {\bibinfo {author} {\bibfnamefont {H.-C.}\ \bibnamefont {Thong}}, \bibinfo {author} {\bibfnamefont {X.}~\bibnamefont {Wang}}, \bibinfo {author} {\bibfnamefont {J.}~\bibnamefont {Han}}, \bibinfo {author} {\bibfnamefont {L.}~\bibnamefont {Zhang}}, \bibinfo {author} {\bibfnamefont {B.}~\bibnamefont {Li}}, \bibinfo {author} {\bibfnamefont {K.}~\bibnamefont {Wang}},\ and\ \bibinfo {author} {\bibfnamefont {B.}~\bibnamefont {Xu}},\ }\bibfield  {title} {\bibinfo {title} {Machine learning interatomic potential for molecular dynamics simulation of the ferroelectric $\mathrm{KNb}{\mathrm{o}}_{3}$ perovskite},\ }\href {https://doi.org/10.1103/PhysRevB.107.014101} {\bibfield  {journal} {\bibinfo  {journal} {Phys. Rev. B}\ }\textbf {\bibinfo {volume} {107}},\ \bibinfo {pages} {014101} (\bibinfo {year} {2023})}\BibitemShut {NoStop}%
\bibitem [{\citenamefont {Feng}\ \emph {et~al.}(2025)\citenamefont {Feng}, \citenamefont {Thong}, \citenamefont {Wang}, \citenamefont {Bi},\ and\ \citenamefont {Xu}}]{Feng-25}%
  \BibitemOpen
  \bibfield  {author} {\bibinfo {author} {\bibfnamefont {N.}~\bibnamefont {Feng}}, \bibinfo {author} {\bibfnamefont {H.-C.}\ \bibnamefont {Thong}}, \bibinfo {author} {\bibfnamefont {K.}~\bibnamefont {Wang}}, \bibinfo {author} {\bibfnamefont {K.}~\bibnamefont {Bi}},\ and\ \bibinfo {author} {\bibfnamefont {B.}~\bibnamefont {Xu}},\ }\bibfield  {title} {\bibinfo {title} {Atomic-scale polarization switching driven by terahertz light in ferroelectric ${\mathrm{knbo}}_{3}$},\ }\href {https://doi.org/10.1103/PhysRevB.111.024305} {\bibfield  {journal} {\bibinfo  {journal} {Phys. Rev. B}\ }\textbf {\bibinfo {volume} {111}},\ \bibinfo {pages} {024305} (\bibinfo {year} {2025})}\BibitemShut {NoStop}%
\bibitem [{\citenamefont {Togo}\ \emph {et~al.}(2023)\citenamefont {Togo}, \citenamefont {Chaput}, \citenamefont {Tadano},\ and\ \citenamefont {Tanaka}}]{Togo-23}%
  \BibitemOpen
  \bibfield  {author} {\bibinfo {author} {\bibfnamefont {A.}~\bibnamefont {Togo}}, \bibinfo {author} {\bibfnamefont {L.}~\bibnamefont {Chaput}}, \bibinfo {author} {\bibfnamefont {T.}~\bibnamefont {Tadano}},\ and\ \bibinfo {author} {\bibfnamefont {I.}~\bibnamefont {Tanaka}},\ }\bibfield  {title} {\bibinfo {title} {Implementation strategies in phonopy and phono3py},\ }\href {https://doi.org/10.1088/1361-648X/acd831} {\bibfield  {journal} {\bibinfo  {journal} {Journal of Physics: Condensed Matter}\ }\textbf {\bibinfo {volume} {35}},\ \bibinfo {pages} {353001} (\bibinfo {year} {2023})}\BibitemShut {NoStop}%
\bibitem [{\citenamefont {Zhang}\ \emph {et~al.}(2024)\citenamefont {Zhang}, \citenamefont {Thong}, \citenamefont {Bastogne}, \citenamefont {Gui}, \citenamefont {He},\ and\ \citenamefont {Ghosez}}]{Zhang-24}%
  \BibitemOpen
  \bibfield  {author} {\bibinfo {author} {\bibfnamefont {H.}~\bibnamefont {Zhang}}, \bibinfo {author} {\bibfnamefont {H.-C.}\ \bibnamefont {Thong}}, \bibinfo {author} {\bibfnamefont {L.}~\bibnamefont {Bastogne}}, \bibinfo {author} {\bibfnamefont {C.}~\bibnamefont {Gui}}, \bibinfo {author} {\bibfnamefont {X.}~\bibnamefont {He}},\ and\ \bibinfo {author} {\bibfnamefont {P.}~\bibnamefont {Ghosez}},\ }\bibfield  {title} {\bibinfo {title} {Finite-temperature properties of the antiferroelectric perovskite ${\mathrm{pbzro}}_{3}$ from a deep-learning interatomic potential},\ }\href {https://doi.org/10.1103/PhysRevB.110.054109} {\bibfield  {journal} {\bibinfo  {journal} {Phys. Rev. B}\ }\textbf {\bibinfo {volume} {110}},\ \bibinfo {pages} {054109} (\bibinfo {year} {2024})}\BibitemShut {NoStop}%
\bibitem [{\citenamefont {Yu}\ \emph {et~al.}(2025)\citenamefont {Yu}, \citenamefont {Anand}, \citenamefont {Zhang}, \citenamefont {Sasani}, \citenamefont {Bastogne}, \citenamefont {Gómez-Ortiz}, \citenamefont {Xu},\ and\ \citenamefont {Philippe}}]{dataverse_BTO_miao}%
  \BibitemOpen
  \bibfield  {author} {\bibinfo {author} {\bibfnamefont {M.}~\bibnamefont {Yu}}, \bibinfo {author} {\bibfnamefont {S.}~\bibnamefont {Anand}}, \bibinfo {author} {\bibfnamefont {J.}~\bibnamefont {Zhang}}, \bibinfo {author} {\bibfnamefont {A.}~\bibnamefont {Sasani}}, \bibinfo {author} {\bibfnamefont {L.}~\bibnamefont {Bastogne}}, \bibinfo {author} {\bibfnamefont {F.}~\bibnamefont {Gómez-Ortiz}}, \bibinfo {author} {\bibfnamefont {H.}~\bibnamefont {Xu}},\ and\ \bibinfo {author} {\bibfnamefont {G.}~\bibnamefont {Philippe}},\ }\href {https://doi.org/10.58119/ULG/KD4XC6} {\bibinfo {title} {{Second-Principles Model of BaTiO3 used on "Role of long-range dipolar interactions in the simulation of the properties of polar crystals using effective atomic potentials"}}} (\bibinfo {year} {2025})\BibitemShut {NoStop}%
\end{thebibliography}
\end{document}